\documentclass[%
 aip,
 jmp,
 amsmath,amssymb,
reprint, onecolumn 
]{revtex4-1}

\usepackage{graphicx}
\usepackage{dcolumn}
\usepackage{bm}

\usepackage[utf8]{inputenc}
\usepackage[T1]{fontenc}
\usepackage{mathptmx}
\usepackage{etoolbox}

\usepackage{cancel}
\usepackage{enumitem}
\usepackage{placeins}

\usepackage{mathtools}
\usepackage{makecell,tabularx}
\setcellgapes{1pt}
\makegapedcells

\makeatletter
\def\@email#1#2{%
 \endgroup
 \patchcmd{\titleblock@produce}
  {\frontmatter@RRAPformat}
  {\frontmatter@RRAPformat{\produce@RRAP{*#1\href{mailto:#2}{#2}}}\frontmatter@RRAPformat}
  {}{}
}%
\makeatother

\DeclareMathOperator{\erf}{erf}
\DeclareMathOperator{\erfc}{erfc}
\begin{document}
\def\opsim{\mathop{\sim}}
\def\opsimeq{\mathop{\simeq}}

\preprint{AIP/123-QED}


\title{Conditioning the \emph{tanh-drift} process and related diffusions on first-passage times: Exact drifts, bridges, and process equivalences}
\author{Kacim Fran\c{c}ois-\'{E}lie}
\affiliation{Universit\'{e} Paris-Saclay, CEA, Service d'\'Etudes des R\'eacteurs et de Math\'ematiques Appliqu\'ees, 91191, Gif-sur-Yvette, France}
\author{Alain Mazzolo}
\affiliation{Universit\'e Paris-Saclay, CEA, Service d'\'Etudes des R\'eacteurs et de Math\'ematiques Appliqu\'ees, 91191, Gif-sur-Yvette, France}
\email{alain.mazzolo@cea.fr}
\email{kacim.francois-elie@cea.fr}

\date{\today}

\begin{abstract}
In this article, we consider the Bene\v{s} process with drift 
$\mu(x)=\alpha \tanh(\alpha x + \beta)$, with $\alpha > 0$, $\beta \in \mathbb{R}$, 
that is, the diffusion defined by the stochastic differential equation 
$dX(t)=\alpha \tanh(\alpha X(t)+\beta)\,dt + dW(t)$, with an absorbing barrier at $x=a$.
After deriving the propagator and key associated quantities---the 
first-passage-time distribution and the survival probability---we then condition 
this process to have various prescribed first-passage-time distributions. 
When the conditioning is imposed at an infinite time horizon, this procedure 
reveals the existence of different processes that share the same 
first-passage-time distribution as the Bene\v{s} process, a phenomenon recently 
observed in the case of Brownian motion with drift.
When the conditioning is imposed at a finite time horizon, the procedure shows 
that the conditioned Bene\v{s} process and the Brownian motion with drift under the 
same conditioning exhibit identical behaviors. This strengthens an elegant result 
of Benjamini and Lee stating that Brownian motion and the Bene\v{s} process share 
the same Brownian bridge, and it also connects with more recent findings obtained 
by conditioning two independent identical Brownian motions with drift, or two independent
Bene\v{s} processes that annihilate upon meeting. Moreover, we show that several conditioned Bene\v{s} drifts converge near the 
absorbing boundary to the drift of the taboo diffusion, i.e., the diffusion conditioned to never reach the absorbing boundary, which motivates a parallel 
analysis of the taboo process itself. Using Girsanov’s theorem, we derive its 
propagator, first-passage-time distribution, and conditioned versions, thereby 
further clarifying the structural relationships between Bene\v{s}, Brownian, 
and taboo dynamics.
\end{abstract}

\maketitle

\newcommand{\E}{\mathrm{E}}
\newcommand{\Var}{\mathrm{Var}} 
\newcommand{\Cov}{\mathrm{Cov}}



\section{Introduction}
\label{sec_intro}
The first-passage-time (FPT) problem---namely, determining the distribution 
of the time at which a stochastic process first reaches a prescribed boundary---is 
a fundamental question in the theory of diffusion processes~\cite{ref_book_Redner,ref_intro_Redner}. It plays a central role 
in a wide range of scientific areas, including probability theory, statistics, physics, 
chemistry, mathematical finance, neuroscience, reliability theory~\cite{ref_book_Metzler} 
and, more recently, biology~\cite{ref_McKenzie}. In particular, first-passage-time problems play a central role in target search processes, where a diffusing particle aims to locate a given target~\cite{ref_book_Grebenkov}. In medicine, FPT models have become increasingly 
important, for instance in oncology, where the time at which a tumor first reaches 
a clinically relevant size is used to assess the efficacy of anticancer therapies and 
to characterize key temporal events in tumor progression~\cite{ref_Otunuga}.

Despite its broad relevance, explicit first-passage-time densities are known only 
for a limited class of one-dimensional diffusion processes. Closed-form expressions are available for the Wiener process~\cite{ref_book_Redner,ref_Monthus_Mazzolo}, 
for the Bessel diffusion $\mathrm{BES}^\delta$ for all dimensions $\delta>0$, 
both for hitting zero (when it is accessible) and for hitting any positive level~\cite{ref_book_Borodin}, 
and for a special case of the Ornstein--Uhlenbeck (OU) process—specifically, when the absorbing boundary is located at the origin~\cite{ref_Alili,ref_Ricciardi}. Beyond these specific settings, obtaining exact FPT 
distributions remains a challenging open problem. Even for the OU process, one of 
the most widely used models in physics and neuroscience, the FPT density for a 
general threshold admits no closed-form expression. These limitations have 
motivated the development of alternative methods, including transformations of 
known solvable processes~\cite{ref_Metzler} and conditioning techniques~\cite{ref_Monthus_Mazzolo,ref_Mazzolo_SAA}, to explore new families of diffusion processes with analytically tractable first-passage properties. Processes conditioned to satisfy constraints also arise in stochastic thermodynamics, 
for instance in the context of work production along trajectories subject to absorption 
constraints~\cite{ref_PRL_2025}. This connection further highlights the potential relevance 
of conditioning approaches beyond purely mathematical considerations.

\smallskip
Building on this perspective, there is considerable interest in identifying diffusions for which first-passage properties can be determined exactly, or in constructing new processes with prescribed first-passage-time statistics. 
In this work, we first focus on the \emph{tanh-drift} process~\cite{ref_Mazzolo_Monthus_annihilation} (also referred to as the Bene\v{s} process~\cite{ref_book_Sarkka}), 
a diffusion $X(t)$ characterized by the drift $\mu(x) = \alpha \tanh(\alpha x + \beta)$, with $\alpha > 0$, $\beta \in \mathbb{R}$,  and governed by the It\^o stochastic differential equation (SDE):
\begin{equation}
\label{eq_diffusion_tanh-drift}
   \left\{
       \begin{aligned}
          dX(t) & =  \alpha \tanh(\alpha X(t) + \beta)\, dt + dW(t), \qquad t \ge 0 \\       
          X(0)  & =  x_0 \, ,
       \end{aligned}
   \right.
\end{equation}
where $W(t)$ is a standard Brownian motion. 
Furthermore, the process is absorbed at a fixed barrier $x = a$, which we take to be positive without loss of generality.

Building on recent advances in conditioning diffusions with respect to their 
first-passage-time distributions, we derive closed-form expressions for the 
drift of the \emph{tanh-drift} process conditioned on various prescribed FPT 
densities, together with explicit propagators for the resulting diffusions. 
This approach not only yields efficient simulation schemes, but also uncovers 
structural correspondences between seemingly different processes. In 
particular, we show that the \emph{tanh-drift} process, Brownian motion with drift, 
and certain inhomogeneous conditioned diffusions may share identical 
first-passage-time distributions under suitable conditioning. Moreover, some 
of the conditioned \emph{tanh-drift} drifts converge near the absorbing boundary to 
the drift of the taboo process. Motivated by this observation, we also extend 
our analysis to the taboo diffusion itself and derive its first-passage-time 
distribution, survival probability, and conditioned versions. This provides an 
additional illustration of the unifying character of the conditioning 
framework and further clarifies the relationships between these families of 
diffusion models.

\smallskip
The article is organized as follows.
Section~\ref{sec_girsanov} recalls Girsanov’s theorem and illustrates it on simple examples before extending it to the \emph{tanh-drift} process, for which we derive the propagator in the presence of an absorbing barrier together with the associated first-passage-time density.
In Section~\ref{sec_Conditioned_tanh-drift}, we develop the main conditioning results for the \emph{tanh-drift} diffusion, considering both finite and infinite time horizons as well as conditioning with respect to various prescribed FPT laws.
Section~\ref{sec_Conditioned_BM} is devoted to the corresponding results for conditioned Brownian motion, while Section~\ref{sec_Conditioned_taboo} extends the conditioning framework to the taboo process and highlights further structural connections with the \emph{tanh-drift} and Brownian cases.
Finally, Section~\ref{sec_Conclusion} presents some concluding remarks.

For clarity, summary tables collecting all the induced drifts associated with the conditioning schemes considered in this work are provided in Appendix~\ref{appendix_tables}, and Appendix~\ref{appendix_inverse} discusses reversibility properties under FPT conditioning.

\section{Girsanov Theorem}
\label{sec_girsanov}

\subsection{Motivations for using Girsanov’s theorem}
The probability density function is a fundamental quantity for characterizing diffusion processes~\cite{ref_book_Karlin,ref_book_Risken,ref_book_Pavliotis}. However, it is rarely known explicitly, except in a few well-established cases, such as Brownian motion, the Ornstein-Uhlenbeck process and their bridges, geometric Brownian motion, Brownian excursions, and Bessel processes~\cite{ref_book_Borodin}. Determining the probability distribution $p(x,t)$ of a diffusion process $X(t)$ governed by the Itô stochastic differential equation (SDE):
\begin{equation}
\label{eq_diffusion}
   \left\{
       \begin{aligned}
	  dX(t) & = \mu(X(t),t) dt + dW(t) \, , ~~ t \geq 0   \\       
	  X(0)  & = x_0 \, ,
       \end{aligned}
   \right.
\end{equation}
where $W(t)$ is a standard Brownian motion, generally requires solving the associated Fokker-Planck equation:
\begin{equation}
\label{eq_fokker_planck_density_1D}
	 \frac{\partial p(x,t)}{\partial t} = -  \frac{\partial \left[\mu(x,t) p(x,t)\right]}{\partial x} + \frac{1}{2} \frac{\partial^2 p(x,t)}{\partial x^2} 
\end{equation}
subject to appropriate boundary conditions. Several analytical techniques are available for solving such partial differential equations, including Fourier~\cite{ref_book_Risken} and Laplace~\cite{ref_Frisch,ref_Mazzolo_Monthus_1,ref_Mazzolo_Monthus_2,ref_Mazzolo_Monthus_3} transforms, the method of characteristics~\cite{ref_Kwok}, eigenfunction expansions~\cite{ref_book_Karlin,ref_book_Risken,ref_book_Pavliotis}, as well as path integral formulations~\cite{ref_book_Risken} and Feynman-Kac technique~\cite{ref_Kac}. Furthermore, in the presence of absorbing or reflecting boundaries, the method of images is particularly effective~\cite{ref_book_Redner,ref_intro_Redner}. However, even with these techniques at our disposal, closed-form expressions for the probability density are only available for a limited number of simple drift functions. 
The objective of this section is to demonstrate that Girsanov’s theorem—though rarely used for this purpose—provides a powerful and convenient tool for deriving exact expressions of probability densities under various boundary conditions~\cite{ref_alain_JMP2024}.
To this end, we first recall Girsanov's theorem and apply it to simple and well-known cases before extending its use to the drift  $\mu(x) = \tanh(\alpha x + \beta)$ of the Bene\v{s} process. For this specific case, we derive the exact expression of the free propagator, as well as its form in the presence of an absorbing barrier.

\subsection{Girsanov’s theorem in practice}
As recalled above, Girsanov’s theorem—whose derivation can be found, for example, in Karatzas and Shreve~\cite{ref_book_Karatzas} or \O ksendal~\cite{ref_book_Oksendal}—is a fundamental result in stochastic analysis that enables the transformation of probability measures for SDEs. In essence, this theorem states that for any diffusion process with drift, one can always transform it into an equivalent drift-free world by applying an appropriate change of measure. In other words, in this drift-free (and thus hopefully simpler) world, the trajectories are weighted by a coefficient that accounts for the original drift, which is precisely given by Girsanov's theorem. When this weight is computable, one can obtain the exact expression of the probability density, as we will now see. To be concrete, let us consider the diffusion defined by its stochastic representation~Eq.\eqref{eq_diffusion}. Girsanov's theorem states that the expectation of any function $h(X(t))$, where $X(t)$ is a solution of Eq.~\eqref{eq_diffusion}, can be expressed as~\cite{ref_book_Sarkka}  
\begin{equation}  
\label{def_new_expectation}  
	E[h(X(t))] = E[Z(t)h(x_0 + W(t))] \, ,  
\end{equation}  
where $\tilde{X}(t) = x_0 + W(t)$ is a driftless process (standard Brownian motion), and the weighting factor $Z(t)$ is given by  
\begin{equation}  
\label{def_Z}  
	Z(t) = e^{\textstyle\int_0^t \mu(x_0 + W(u),u) dW(u) - \frac{1}{2} \int_0^t \mu(x_0 + W(u),u)^2 du} \, .  
\end{equation}
Observe that the first integral in the exponential is a stochastic integral, which is often challenging to compute in practice.



\subsection{First example: from a standard Brownian motion to a Brownian motion with constant drift}
\label{subsec_girsanov_1}
We start by examining the simplest case, namely a diffusion with a constant drift, and seek to derive various probability densities for this drifted diffusion based on those of standard Brownian motion.
In the case of a constant drift $\mu(x,t) = \mu$, and the weight $Z(t)$ is given by
\begin{equation} 
\label{def_Z_mu_constant}
	Z(t) = e^{\mu (W(t) -x_0) - \frac{1}{2} \mu^2 t}  \, .
\end{equation}  
Since this expression depends only on the state of the Brownian motion at time $t$, the probability density can be explicitly determined~\cite{ref_book_Sarkka}. Given that the probability density of the  process $\tilde{X}(t) = x_0 + W(t) $ is 
\begin{equation}
\label{density_driftless}
	 p(\tilde{X},t)  =  \frac{1}{\sqrt{2 \pi t}} e^{- \frac{(\tilde{X} -x_0)^2}{2t}},~t>0,
\end{equation}
from the preceding equation and Eq.\eqref{def_Z_mu_constant} we obtain the probability density $p_{\mu}(x,t)$ of the Brownian with constant drift, that is
\begin{equation}
	 p_{\mu}(x,t)  = e^{\mu (x - x_0) - \frac{1}{2} \mu^2 t}\times \frac{1}{\sqrt{2 \pi t}} e^{- \frac{(x -x_0)^2}{2t}} ,~t>0 .
\end{equation}
By rearranging the terms, we recover the well-known expression for the propagator of Brownian motion with a constant drift $\mu$, namely:
\begin{equation}
	 p_{\mu}(x,t)  =  \frac{1}{\sqrt{2 \pi t}} e^{- \frac{((x - x_0) - \mu t)^2}{2t}} ,~t>0 .
\end{equation}
Of course, one might reasonably think that using Girsanov's theorem to obtain this result is an overly complicated approach. However, once the weight $Z(t)$ is computed, the flexibility of this method allows for solving more complex cases with the same ease, as we will demonstrate in the following examples.

\subsection{Second example: from a Brownian motion with an absorbing barrier at $a$ to a Brownian motion with drift and the same absorbing barrier.}
\label{subsec_girsanov_2}
Compared to the previous case, only the geometry has changed; consequently, $Z(t)$ remains the same. Given that the propagator of the Brownian motion with absorbing condition at position $a>0$ is 
\begin{equation}
\label{density_BM_abs}
	 p_{\mathrm{a}}(x,t)  =  \frac{1}{\sqrt{2 \pi t}} \left\{e^{-\frac{(x - x_0)^2}{2t}} - e^{-\frac{(x + x_0 - 2 a)^2}{2t}} \right\},~t>0 ,
\end{equation}
we can immediately deduce, using Girsanov's theorem, that the propagator of the Brownian motion with drift $\mu$ and absorbing condition at position $a>0$ is
\begin{equation}
\label{density_BM_abs_Girsanov}
	 p_{\mathrm{a}}^{\mu}(x,t)  =  e^{\mu (x - x_0) - \frac{1}{2} \mu^2 t} \times \frac{1}{\sqrt{2 \pi t}} \left\{e^{-\frac{(x - x_0)^2}{2t}} - e^{-\frac{(x + x_0 - 2 a)^2}{2t}} \right\},~t>0 .
\end{equation}
Again, by rearranging the terms, we recover the well-known propagator of Brownian motion with drift $\mu$ with an absorbing barrier at $a$, 
\begin{equation}
\label{density_BM_abs_images}
	 p_{\mathrm{a}}^{\mu}(x,t)  =   \frac{1}{\sqrt{2 \pi t}} \left\{e^{-\frac{(x - x_0 -\mu t)^2}{2t}} - e^{2 \mu (a-x_0)} e^{-\frac{(x + x_0 - 2 a -\mu t)^2}{2t}} \right\},~t>0 ,
\end{equation}
which is usually derived using the method of images\cite{ref_book_Redner,ref_Monthus_Mazzolo}.

\subsection{The \emph{tanh-drift} process}
\label{subsec_girsanov_tanh-drift}
The \emph{tanh-drift} process~\cite{ref_Monthus_Mazzolo}, also known as the Bene\v{s} process~\cite{ref_book_Sarkka}, or referred to as the hyperbolic Ornstein--Uhlenbeck process~\cite{ref_Borodin} in the mathematical literature, is a diffusion process characterized by a space-dependent drift of the form
\begin{equation}
    \mu(x) = \alpha \tanh(\alpha x + \beta) \, ,
\end{equation}
and, since the hyperbolic tangent function is odd, we may assume without loss of generality that $\alpha > 0$. 
Note that the drift is repulsive, driving the process away from the origin. As in the case of the Brownian motion with drift, we first derive the propagator of the \emph{tanh-drift} process and then the propagator of the \emph{tanh-drift} process with an absorbing barrier at $x=a$. To this aim, we first evaluate the weighting factor $Z(t)$ [Eq.\eqref{def_Z}] which worth in this case   
\begin{equation}  
\label{def_Z_tanh-drift}  
	Z(t) = e^{\textstyle\int_0^t \alpha  \tanh (\alpha (x_0 + W(u)) + \beta) dW(u) - \frac{1}{2} \int_0^t   \alpha^2  \tanh^2 (\alpha (x_0 + W(u)) + \beta)  du} \, .  
\end{equation}
Applying Itô's formula to the function $\log (\cos(\alpha x + \beta))$
\begin{equation}
 d\log\!\big(\cosh(\alpha W(t) + \beta)\big)
= \frac{1}{2}\,\alpha^{2}  \frac{dt}{{\cosh}^{2}(\alpha W(t) + \beta)}
+ \alpha\,\tanh(\alpha W(t) + \beta)\,dW(t) 
\end{equation}
allows us to express $Z(t)$ as
\begin{align}
\label{def_Z_tanh-drift_final}
	Z(t) & =  e^{{\big{[}} \textstyle \log\!\big(\cosh(\alpha (x_0 + W(u)) + \beta)\big) {\big{]}}_0^t - \frac{1}{2} \alpha^2 \textstyle \int_0^t \overbrace{\left\{ \frac{1}{{{\cosh}^{2} (\alpha (x_0 + W(u)) + \beta) }} + \tanh^2 (\alpha (x_0 + W(u)) + \beta) \right\} }^{= 1} du} 
	 \nonumber \\
			 & = \frac{\cosh(\alpha (x_0 + W(t)) + \beta)}{\cosh(\alpha x_0 + \beta)} e^{- \frac{1}{2} \alpha^2 t} \, .
\end{align}
Since $Z(t)$ depends solely on the state of the Brownian motion at time $t$, the transition density of the \emph{tanh-drift} process can be determined in closed form and is given by
\begin{equation}
\label{density_tanh-drift}
	 	 p(x,t)  = \frac{\cosh(\alpha x + \beta)}{\cosh(\alpha x_0 + \beta)} e^{- \frac{1}{2} \alpha^2 t}  \times \frac{1}{\sqrt{2 \pi t}} e^{- \frac{(x -x_0)^2}{2t}} ,~t>0 .
\end{equation}
An expression that can be found in~\cite{ref_Borodin}, where it is derived by means of Green’s function analysis and Laplace transform techniques. Similarly, the expression of $Z(t)$ allows one to derive the propagator of the \emph{tanh-drift} process with an absorbing boundary at $x=a$. We get
\begin{equation}
\label{density_tanh-drift_abs}
	 	 p_a(x,t)  = \frac{\cosh(\alpha x + \beta)}{\cosh(\alpha x_0 + \beta)} e^{- \frac{1}{2} \alpha^2 t}  \times \frac{1}{\sqrt{2 \pi t}} \left(e^{-\frac{(x - x_0)^2}{2t}} - e^{-\frac{(x + x_0 - 2 a)^2}{2t}} \right),~t>0 .
\end{equation}
As the above relation holds for arbitrary initial conditions, it follows that the transition density takes the form
\begin{equation}
\label{density_tanh-drift_abs_full}
	 	 p_a(x_2,t_2|x_1,t_1)  = \frac{\cosh(\alpha x_2 + \beta)}{\cosh(\alpha x_1 + \beta)} \frac{e^{- \frac{1}{2} \alpha^2 (t_2 - t_1)}} {\sqrt{2 \pi (t_2 - t_1)}} \bigg(e^{-\frac{(x_2 - x_1)^2}{2 (t_2 - t_1)}} - e^{-\frac{(x_2 + x_1 - 2 a)^2}{2(t_2 - t_1)}} \bigg) ,~t_2>t_1.
\end{equation}
Once the propagator is known, it becomes straightforward to determine quantities that characterize the processes, such as the first-passage time density or the survival probability, which will be essential when conditioning the process.

The probability density  $\gamma(t_2 \vert x_1,t_1)$ of the first-passage time $t_2$ to the barrier $a$ can be obtained from the spatial derivative of the propagator in Eq.~(\ref{density_tanh-drift_abs}) with respect to $x_2$~\cite{ref_Monthus_Mazzolo}
\begin{equation}
\label{gammafirstpassagetanh-drift}
   \gamma(t_2 \vert x_1,t_1) = - \frac{1 }{2}  \left( \partial_{x_2} p_a(x_2,t_2 \vert x_1,t_1) \right)\bigg\vert_{x_2=a} =  \frac{(a-x_1) }{\sqrt{2 \pi (t_2-t_1)^3} } \frac{\cosh(\alpha a + \beta)}{\cosh(\alpha x_1 + \beta)} e^{-\frac{(a-x_1)^2+\alpha ^2 (t_2-t_1)^2}{2 (t_2-t_1)}} .
\end{equation}
Integrating this quantity over all possible times $t_2 \in [t_1, +\infty)$ yields the probability that the process is eventually absorbed 
\begin{equation}
\label{prob_absorption_tanh-drift}
   \int_{t_1}^{\infty} \gamma(t_2 \vert x_1,t_1) dt_2 =  \frac{\cosh(\alpha a + \beta)}{\cosh(\alpha x_1 + \beta)} e^{-(a-x_1) \alpha }  ,
\end{equation}
and the complementary quantity $1 - \int_{t_1}^{\infty} \gamma(t_2 \vert x_1,t_1) \, dt_2$ corresponds to the forever-survival probability $ S(\infty \vert x_1)$ of the process,
\begin{equation}
\label{prob_survival_tanh-drift}
   S(\infty \vert x_1) = 1 - \frac{\cosh(\alpha a + \beta)}{\cosh(\alpha x_1 + \beta)} e^{-(a-x_1) \alpha }  .
\end{equation}
Observe that for $\alpha \neq 0$, the survival probability is always strictly greater than zero, implying that the process can survive indefinitely. This behavior originates from the form of the drift: when the process wanders far into negative values, since $\lim_{x \to -\infty} \mu(x) = - \alpha $, it experiences a negative drift that tends to push it away from the absorbing barrier $a$. In the same vein, the survival probability of the process up to time $T$ can be expressed in terms of the complementary Error function $\erfc(x) = 1 - \erf(x) = 1 - \frac{2}{\sqrt{\pi}} \int_0^{x} du \, e^{- u^2}$ as 
\begin{eqnarray}
\label{prob_survival_tanh-drift_up_to_T}
   S(T \vert x_1,t_1) & = & 1 -  \int_{t_1}^{T} \gamma(t_2 \vert x_1,t_1) \, dt_2 \nonumber \\
                      & = & 1 - \frac{1}{2} \frac{\cosh(\alpha a + \beta)}{\cosh(\alpha x_1 + \beta)} \left(e^{(a-x_1)\alpha} \text{erfc}\left(\frac{a - x_1 +\alpha (T-t_1)}{\sqrt{2(T-t_1)}}\right)+  e^{-(a-x_1) \alpha } \text{erfc} \left( \frac{a-x_1 -\alpha (T-t_1)}{\sqrt{2 (T - t_1)}} \right)   \right)  \, ,
\end{eqnarray}
an expression that can also be obtained by integrating the probability density function $p_a(x_2, T \mid x_1, t_1)$ [Eq.~(\ref{density_tanh-drift_abs})] over the spatial variable $x_2 \in (-\infty, a]$.

\section{Conditioning the \emph{tanh-drift} process}
\label{sec_Conditioned_tanh-drift}
With the results derived in the preceding section, we are now ready to condition the \emph{tanh-drift} process on various first-passage time densities, corresponding to full or partial survival. Before proceeding, we briefly recall the procedure for conditioning a diffusion with respect to its first-passage time density. Since the first-passage time is itself a random time, this type of conditioning is a special case of conditioning with respect to a random time, whose general theory is still missing. To the best of our knowledge, the earliest contribution to this field is due to Baudoin~\cite{ref_Baudoin}, and was later pursued in~\cite{ref_Larmier}. However, it is only very recently that conditioning with respect to the first-passage time and the survival probability at time $t$--- i.e., the probability that a particle has not been absorbed at the boundary up to time $t$ --- has been established~\cite{ref_Monthus_Mazzolo}. For a precise formulation of conditioning with respect to a probability density and an event, we refer the reader to this reference.

Let us recall that, starting from an unconditioned process $X(t)$ with drift $\mu(x)$ satisfying the stochastic differential equation
\begin{equation}
\label{eq_diffusion_unconditioned}
	  dX(t)  = \mu(X(t),t) dt + dW(t) \, , ~~ t \geq 0       ,        
\end{equation}
the conditioned process $X^*(t)$, associated with a prescribed first-passage time probability density (which may be singular, continuous, normalized or not), obeys the SDE
\begin{equation}
\label{eq_diffusion_conditioned}
	  dX^*(t) = \mu^*(X^*(t),t) dt + dW(t) \, , ~~ t \geq 0       ,
\end{equation}
where the conditioned drift is given by
\begin{equation}
	\mu^*(x,t)  = \mu(x) + \partial_x \log  Q(x,t) \, ,
\end{equation}
with the function $Q(x,t)$ encapsulating all the information related to the conditioning.
For the different conditioning schemes, the corresponding expressions of $Q(x,t)$
have been derived in Ref.~\cite{ref_Monthus_Mazzolo} and will be recalled when needed.

\subsection{Conditioning on a finite time horizon $T$}
\label{subsec_conditoning_finite_horizon}
When conditioning the process to have a specified probability distribution $P^*(y,T)$ for being at position $y$ at time $T$, together with a prescribed first-passage-time distribution $\gamma^*(T_a)$ for $T_a \in [0,T]$ (these two quantities are not independent but are related through the survival probability $S^*(t)$ at time $t$: $S^*(T ) = \int_{-\infty}^a P^*(y,T ) dy = 1- \int_{0}^T \gamma^*(T_a )dT_a )$, the conditioned drift is then given by:
\begin{eqnarray}
\label{driftfiniteT}
    \mu^*_T(x,t)  = \mu(x) + \partial_x \log Q_T(x,t) =  \alpha  \tanh (\alpha  x + \beta) + \partial_x \log Q_T(x,t) 
\end{eqnarray}
where $Q_T(x,t)$ is given by [Eq.(34) of Ref.~\cite{ref_Monthus_Mazzolo}]:
\begin{eqnarray}
 \label{QsiniteT}
    Q_T(x,t)  = \int_t^{T} \gamma^*(T_a) \frac{\gamma(T_a \vert x,t)}{\gamma(T_a\vert 0,0)}  dT_a 
+   \int_{-\infty}^a P^*(y,T )  \frac{ p_a(y,T \vert x,t)  }{p_a(y,T \vert 0,0) } dy  \, .
\end{eqnarray}
The ratio of the first-passage-time densities is obtained from Eq.(\ref{gammafirstpassagetanh-drift}):
\begin{eqnarray}
 \label{ratiogamma}
 \frac{\gamma(T_a \vert x,t)}{\gamma(T_a\vert 0,0)} = \frac{(a-x)}{a} \frac{\cosh(\beta)}{\cosh(\alpha x + \beta) } \frac{T_a^{3/2}}{(T_a-t)^{3/2}}  e^{-\frac{(a-x)^{2}+\alpha^{2}(T_a-t)^{2}}{2(T_a-t)} + 
\frac{a^{2}+\alpha^{2} T_a^{2}}{2T_a} } \, ,
\end{eqnarray}
and the ratio of the propagators provided by Eq.(\ref{density_tanh-drift_abs_full}):
\begin{eqnarray}
 \label{ratiop_a}
   \frac{ p_a(y,T \vert x,t)  }{p_a(y,T \vert 0,0) } = \frac{\cosh(\beta)}{\cosh(\alpha x + \beta) } \frac{\sqrt{T}}{\sqrt{T-t}} \frac{e^{\frac{\alpha^{2} t}{2}}
\left(e^{-\frac{(x-y)^{2}}{2 (T-t)}}-e^{-\frac{(2 a-x-y)^{2}}{2 (T-t)}}\right)}
{\left(e^{-\frac{y^{2}}{2 T}}-e^{-\frac{(y-2 a)^{2}}{2 T}}\right)}
\, .
\end{eqnarray}
Substituting these two quantities into the expression for $Q_T(x,t)$, we obtain:
\begin{eqnarray}
 \label{QsiniteTsimplified}
    Q_T(x,t) = \frac{\cosh(\beta) e^{\frac{\alpha^2 t}{2}}}{\cosh(\alpha x + \beta) } \bigg[ \displaystyle  & & \frac{(a-x)}{a} \int_t^{T} \gamma^*(T_a) \sqrt{\frac{T_a^3}{(T_a-t)^3}} e^{-\frac{(a-x)^{2}}{2(T_a-t)} + \frac{a^{2}}{2 T_a}} dT_a   
    \nonumber \\    
    \displaystyle  & & + 
    \sqrt{\frac{T}{T-t}} \int_{-\infty}^a P^*(y,T )  \frac{
\left(e^{-\frac{(x-y)^{2}}{2 (T-t)}}-e^{-\frac{(2 a-x-y)^{2}}{2 (T-t)}}\right)}
{\left(e^{-\frac{y^{2}}{2 T}}-e^{-\frac{(y-2 a)^{2}}{2 T}}\right)} dy \bigg] \, .
\end{eqnarray}
The expression for the conditioned drift becomes [Eq.(\ref{driftfiniteT})]:
\begin{eqnarray}
\label{driftfiniteTsimplified}
    \mu^*_T(x,t)  = \partial_x \log \left[ \displaystyle  \frac{(a-x)}{a}  \int_t^{T} \gamma^*(T_a) \sqrt{\frac{T_a^3}{(T_a-t)^3}} e^{-\frac{(a-x)^{2}}{2(T_a-t)} + \frac{a^{2}}{2 T_a}} dT_a   \displaystyle  + 
    \sqrt{\frac{T}{T-t}}  \int_{-\infty}^a P^*(y,T )  \frac{
\left(e^{-\frac{(x-y)^{2}}{2 (T-t)}}-e^{-\frac{(2 a-x-y)^{2}}{2 (T-t)}}\right)}
{\left(e^{-\frac{y^{2}}{2 T}}-e^{-\frac{(y-2 a)^{2}}{2 T}}\right)} dy  \right] \, ,
\end{eqnarray}
which is independent of the initial drift $\mu(x) = \alpha \tanh(\alpha x + \beta)$. 
The fact that conditioned processes can be independent of the original drift has also been observed for Brownian motion with constant drift~\cite{ref_Monthus_Mazzolo}, and has attracted significant interest in recent years~\cite{ref_Krapivsky,ref_Larmier}. We refer the reader to these references for detailed discussions.
Let us emphasize the important fact that the expression [Eq.~(\ref{driftfiniteTsimplified})] obtained when conditioning the \emph{tanh-drift} process on a finite time horizon $T$ is identical to that obtained when conditioning a Brownian motion with constant drift under the same hypothesis [Eqs.~(69) and (70) of Ref.~\cite{ref_Monthus_Mazzolo}]. As a consequence, all the results derived in Ref.~\cite{ref_Monthus_Mazzolo} for the Brownian motion with constant drift conditioned on a finite time horizon $T$ can be directly transposed to the \emph{tanh-drift} process. For instance, in the particular case where the process is conditioned to be fully absorbed at a single prescribed time $T^* \in (0,T)$, the first-passage-time density reduces to a Dirac delta function, $\gamma^*(T_a) = \delta(T_a - T^*)$, and the conditioned drift is given by [Eq.~(76) of Ref.~\cite{ref_Monthus_Mazzolo}]:
\begin{eqnarray}
\mu^*(x,t) = -\frac{1}{a-x} +  \frac{a-x}{T^*-t} \, .
\label{driftdoobdeltaT}
\end{eqnarray}
As observed in Ref.~\cite{ref_Monthus_Mazzolo}, when the level $a$ becomes large ($a \to \infty$), the first term on the right-hand side of Eq.~(\ref{driftdoobdeltaT}) becomes negligible compared to the second one, except when $x$ approaches $a$ near the final time $T^*$. In this case, the drift in Eq.~(\ref{driftdoobdeltaT}) reduces to
\begin{equation}
    \mu^{[BB]}(x,t) = \frac{a - x}{T^* - t},
\label{driftBrownianbridge}
\end{equation}
which is precisely the drift of a Brownian bridge ending at $a$ at the final time $T^*$~\cite{ref_book_Karlin,ref_Majumdar_Orland,ref_Mazzolo_BB}. The fact that Brownian motion with drift and the \emph{tanh-drift} process share the same bridge (namely, the standard Brownian bridge) is not surprising. Indeed, this result appears in Ref.~\cite{ref_Benjamini}, where Benjamini and Lee obtain an even stronger conclusion: the \emph{tanh-drift} process is the only process, together with Brownian motion with drift, that shares this bridge.

\subsection{Conditioning on full survival as $t \to \infty$}
\label{sec_Conditioned_tanh-drift_full_survival}
When the process is conditioned to survive indefinitely, since the forever-survival probability $S(\infty \vert .)$ of the \emph{tanh-drift} process is finite [Eq.(\ref{prob_survival_tanh-drift})], the function $Q(x,t)$  takes the simple form~\cite{ref_Monthus_Mazzolo}
\begin{eqnarray}
 \label{Qspacesurvivalinfinity_def}
    Q_{\infty}^{[\mathrm{surviving}]}(x,t) = \frac{S(\infty \vert x)}{S(\infty \vert x_0)} \, .
\end{eqnarray}
From now on, without loss of generality, we assume that the process starts from $x_0 = 0$ at $t = 0$. Substituting Eq.(\ref{prob_survival_tanh-drift}) into the preceding equation yields:
\begin{eqnarray}
 \label{Qspacesurvivalinfinity_tanhdrift}
    Q_{\infty}^{[\mathrm{surviving}]}(x,t) = \frac{\sinh (\alpha  (a-x))}{\sinh (a \alpha )}  \frac{ \cosh (\beta )}{\cosh (\alpha x + \beta)} \, ,
\end{eqnarray}
and the corresponding conditioned drift reads
\begin{eqnarray}
\mu^*_{\infty}(x,t)  &=& \mu(x) +  \partial_x  \log Q_{\infty}^{[\mathrm{surviving}]}(x,t) 
  = \alpha  \tanh (\alpha  x + \beta)  +  \partial_x \log \left( \frac{\sinh (\alpha  (a-x))}{\cosh (\alpha x + \beta)}  \right)
\nonumber \\
&=& -\alpha \coth(\alpha (a-x))  \, .
\label{driftdoobforever}
\end{eqnarray}
Historically, this drift first appeared in the work of Williams~\cite{ref_Williams}. More importantly, it corresponds to the drift obtained when conditioning a Brownian motion with a negative drift and an absorbing barrier at $x = a > 0$ to survive forever. Consequently, the \emph{tanh-drift} process and the drifted negative Brownian motion share the same conditioned process when conditioned on eternal survival. As we shall see later, these processes also exhibit other remarkable common properties once conditioned.

\subsection{Conditioning on a prescribed first-passage time distribution}
\label{sec_Conditioned_tanh-drift_fpt}
\subsubsection{Conditioning with respect to the first-passage time density $\gamma^*(t)  =  \frac{a }{\sqrt{2 \pi t^3} } e^{- \frac{(a-\mu t )^2}{2 t}} $}
We begin by conditioning the process on a prescribed first-passage time density through the barrier $a$, namely that of a Brownian motion with drift $\mu$. This density is well known and expressed as~\cite{ref_book_Redner,ref_Monthus_Mazzolo}:
\begin{eqnarray}
\gamma^{*}_{\mu}(t_2 \vert x_1,t_1) =  \frac{(a-x_1) }{\sqrt{2 \pi (t_2-t_1)^3}} e^{- \frac{(a-x_1-\mu(t_2-t_1) )^2}{2(t_2-t_1)}} \, ,
\label{gammafirstBrown}
\end{eqnarray}
and the normalization of $\gamma^{*}_{\mu}(t_2 \vert x_1, t_1)$ over all possible finite times $t_2 \in [t_1, +\infty)$ is given by:
\begin{eqnarray}
    \int_{t_1}^{+\infty} \gamma^{*}_{\mu}(t_2 \vert x_1, t_1) d t_2   
=  \left\lbrace
  \begin{array}{lll}
    1  
    &~~\mathrm{if~~} \mu \ge 0
    \\
    e^{2 \mu (a-x_1) } 
    &~~\mathrm{if~~} \mu < 0  
  \end{array}
\right.  \, ,
\label{gammafirstBrowninte}
\end{eqnarray}
and the forever-survival probability $ S^{*}_{\mu}(\infty \vert x_1)$ vanishes only for positive drift $\mu \geq 0$
\begin{eqnarray}
    S^{*}_{\mu}(\infty \vert x_1) = 1 - \int_{t_1}^{+\infty}   \gamma^{*}_{\mu}(t_2 \vert x_1, t_1) d t_2   = 
 \left\lbrace
  \begin{array}{lll}
    0  
    &~~\mathrm{if~~} \mu \ge 0
    \\
    1- e^{2 \mu (a-x_1) } 
    &~~\mathrm{if~~} \mu < 0  
  \end{array}
\right.  \, .
\label{survivalBrown}
\end{eqnarray}
Therefore, when conditioning on the first-passage time density of a Brownian motion with drift, one should carefully distinguish between two cases: $\mu \geq 0$ (full absorption) and $\mu < 0$ (partial absorption).

\bigskip
\paragraph*{Case $\mu \geq 0$: Vanishing survival in the infinite-time limit}\mbox{}\\
When the conditioning is toward a first-passage-time distribution $\gamma^*(T_a)$ that is normalized to unity, the function $Q_{\infty}^{[\mathrm{time}]}(x,t)$ takes the form:
\begin{eqnarray}
    Q_{\infty}^{[time]}(x,t) = 
    \int_t^{+\infty} \gamma^*(T_a) \frac{\gamma(T_a \vert x,t)}{\gamma(T_a\vert 0,0)}  dT_a \, .
 \label{Qtimeinfinity}
\end{eqnarray}
Inserting the expressions of $\gamma^*(T_a) = \gamma^{*}_{\mu}(T_a \vert 0, 0)$ [Eq.~(\ref{gammafirstBrown})] and $\gamma(t_2 \mid x_1, t_1)$ [Eq.~(\ref{gammafirstpassagetanh-drift})] into the preceding equation leads to:
\begin{eqnarray}
    Q_{\infty}^{[time]}(x,t)  = (a-x) \frac{\cosh (\beta )} {\cosh(\beta +\alpha  x)}
    \int_t^{+\infty} \frac{e^{\frac{1}{2} \left(2 a \mu -\frac{(a-x)^2}{T_a-t}+\alpha ^2 t-\mu ^2 T_a\right)}}{\sqrt{2 \pi (T_a-t)^3}} dT_a  =  \frac{\cosh (\beta )} {\cosh(\beta +\alpha  x)} e^{\frac{1}{2} t \left(\alpha ^2-\mu ^2\right)+\mu  x}  \, ,
\end{eqnarray}
and the corresponding conditioned drift reduces to:
\begin{eqnarray}
\mu^*_{\infty}(x,t)  &=& \mu(x) +  \partial_x  \log Q_{\infty}^{[time]}(x,t) 
  = \alpha  \tanh (\alpha  x + \beta)  +  \partial_x \log \left( \frac{\cosh (\beta )} {\cosh(\beta +\alpha  x)} e^{\frac{1}{2} t \left(\alpha ^2-\mu ^2\right)+\mu  x}  \right)
\nonumber \\
&=& \mu  \, .
\label{driftdoobforevermupositive}
\end{eqnarray}
Thus, conditioning on the normalized first-passage time distribution of a Brownian motion with a uniform positive drift $\mu$ simply yields a Brownian motion with drift $\mu$, independently of the initial drift $\mu(x) = \alpha \tanh(\alpha x + \beta)$ of the original \emph{tanh-drift} process. 

\medskip
\paragraph*{Case $\mu < 0$: Partial survival in the infinite-time limit}\mbox{}\\
When conditioning on a first-passage-time distribution $\gamma^*(T_a)$ that is not normalized to unity, the function $Q_{\infty}^{[\mathrm{partial}]}(x,t)$ takes a more involved form and reads, when the forever-survival probability $S(\infty \mid \cdot)$ of the unconditioned process is finite: 
\begin{eqnarray}
 \label{Qinfinitypartial}
    Q^{[\mathrm{partial}]}_{\infty}(x,t)  = 
    \int_t^{\infty} \gamma^*(T_a) \frac{\gamma(T_a \vert x,t)}{\gamma(T_a\vert 0,0)}  dT_a  
    +  S^*(\infty ) \frac{S(\infty \vert x)}{S(\infty \vert 0)}  \, .
\end{eqnarray}
The first term on the right-hand side of the previous equality is an integral that encapsulates the information required to construct a new (non-normalized) first-passage-time density, whereas the second term accounts, in some sense, for the missing contribution from the trajectories that do not reach the barrier, through the survival probabilities. As in the case $\mu \ge 0$, the integral over the first-passage-time densities can be computed explicitly; for $\mu < 0$, one obtains:
\begin{eqnarray} 
    \int_t^{+\infty} \gamma^*(T_a) \frac{\gamma(T_a \vert x,t)}{\gamma(T_a\vert 0,0)}  dT_a =  \frac{(a-x) \cosh (\beta )} {\cosh(\beta +\alpha  x)}
    \int_t^{+\infty} \frac{e^{\frac{1}{2} \left(2 a \mu -\frac{(a-x)^2}{T_a-t}+\alpha ^2 t-\mu ^2 T_a\right)}}{\sqrt{2 \pi (T_a-t)^3}} dT_a  =  \frac{\cosh (\beta )} {\cosh(\beta +\alpha  x)}  e^{2 a \mu +\frac{1}{2} t \left(\alpha ^2-\mu ^2\right)-\mu  x} \, .
\end{eqnarray}
The second quantity, $S^*(\infty)\, S(\infty \mid x) / S(\infty \mid 0)$, follows from Eqs.~(\ref{survivalBrown}) and~(\ref{Qspacesurvivalinfinity_tanhdrift}), and is given by:
\begin{eqnarray}
     S^*(\infty ) \frac{S(\infty \vert x)}{S(\infty \vert 0)} = \left( 1- e^{2 \mu a } \right) \frac{\sinh (\alpha  (a-x))}{\sinh (a \alpha )}  \frac{ \cosh (\beta )}{\cosh (\alpha x + \beta)} \, .
\end{eqnarray}
By combining these two results, one finds:
\begin{eqnarray}
 \label{Qinfinitypartialfinal}
    Q^{[\mathrm{partial}]}_{\infty}(x,t)  = \frac{\cosh (\beta )} {\cosh(\alpha  x + \beta)} \left[ e^{2 a \mu +\frac{1}{2} t \left(\alpha ^2-\mu ^2\right)-\mu  x} + \left( 1- e^{2 \mu a } \right) \frac{\sinh (\alpha  (a-x))}{\sinh (a \alpha )} \right] ,
\end{eqnarray}
and the corresponding conditioned drift takes the form:
\begin{eqnarray}
\mu^*_{\infty}(x,t)  &=& \mu(x) +  \partial_x  \log Q^{[\mathrm{partial}]}_{\infty}(x,t) \nonumber \\
 & = &\alpha  \tanh (\alpha  x + \beta)  +  \partial_x \log \left( \frac{\cosh (\beta )} {\cosh(\alpha  x + \beta)} \left[ e^{2 a \mu +\frac{1}{2} t \left(\alpha ^2-\mu ^2\right)-\mu  x} + \left( 1- e^{2 \mu a } \right) \frac{\sinh (\alpha  (a-x))}{\sinh (a \alpha )} \right]  \right)
\nonumber \\
&=& \displaystyle{\frac{\frac{\alpha  \left(e^{2 a \mu }-1\right) \left(e^{2 \alpha  (a-x)}+1\right) e^{x (\alpha +\mu )}}{e^{2 a \alpha }-1}-\mu  e^{2 a \mu +\frac{1}{2} (\alpha^2 -\mu^2 ) t}}{e^{2 a \mu +\frac{1}{2} (\alpha^2 -\mu^2) t}-\frac{\left(e^{2 a \mu }-1\right) \left(e^{2 \alpha  (a-x)}-1\right) e^{x (\alpha +\mu )}}{e^{2 a \alpha }-1}}  }\, .
\label{driftdoobforevermunegative}
\end{eqnarray}
First, observe that this conditioned drift no longer depends on $\beta$. The fact that the conditioned drift is independent of $\beta$ should not be viewed as a direct consequence of the change of variable $y=x+\beta/\alpha$ alone, since this transformation also shifts the absorbing barrier and the initial position. Rather, in the present case, the relevant quantities entering the conditioning procedure depend only on translation-invariant combinations, such as the distance to the barrier, or exhibit an explicit cancellation of the $\beta$-dependence. More importantly, by factoring out $\alpha$, the expression can be rewritten in the form:
\begin{eqnarray}
\label{driftdoobmuneglambda}
    \mu^*_{\infty}(x,t)   = \alpha + \frac{-(\mu +\alpha ) e^{2 a \mu -\mu  x-\alpha  x +\frac{1}{2} \left(\alpha ^2-\mu ^2\right)t}+2 \alpha \frac{\left(1-e^{2 a \mu }\right)}{\left(1-e^{2 a \alpha }\right)} e^{2 \alpha (a-x)}}{e^{2 a \mu -\mu  x-\alpha  x +\frac{1}{2} \left(\alpha ^2-\mu ^2\right)t} +\frac{\left(1-e^{2 a \mu }\right)}{\left(1-e^{2 a \alpha }\right)} \left(1-e^{2 \alpha (a-x)}\right) } \, .
\end{eqnarray}
This is precisely the conditioned drift derived in Ref.~\cite{ref_Mazzolo_SAA} (referred to therein as the type-II process), obtained when conditioning a Brownian motion with a constant negative drift $\alpha < 0$ on a target first-passage-time density $\gamma^{*}_{\mu}(t \vert 0,0)$, which is given by Eq.~(\ref{gammafirstBrown}) with $\mu < 0$.
Thus, the Brownian motion with a constant negative drift and the \emph{tanh-drift} process not only share the same conditioned process when conditioned to survive forever, but also in the case where the conditioning is on an inverse Gaussian density $\gamma^{*}(t) = \frac{a}{\sqrt{2\pi t^3}}\, e^{-\frac{(a - \mu t)^2}{2t}}$, regardless of whether $\mu$ is positive—corresponding to full absorption—or negative—corresponding to partial absorption. The process $X^*(t)$, associated with this drift, and therefore governed by the stochastic differential equation 
\begin{equation}
\label{eq_diffusion_conditioned_typeII}
	  dX^*(t) = \left[ \alpha + \frac{-(\mu +\alpha ) e^{2 a \mu -\mu  X^*(t)-\alpha  X^*(t) +\frac{1}{2} \left(\alpha ^2-\mu ^2\right)t}+2 \alpha \frac{\left(1-e^{2 a \mu }\right)}{\left(1-e^{2 a \alpha }\right)} e^{2 \alpha (a-X^*(t))}}{e^{2 a \mu -\mu  X^*(t)-\alpha  X^*(t) +\frac{1}{2} \left(\alpha ^2-\mu ^2\right)t} +\frac{\left(1-e^{2 a \mu }\right)}{\left(1-e^{2 a \alpha }\right)} \left(1-e^{2 \alpha (a-X^*(t))}\right) } \right] dt + dW(t) \, , ~~ t \geq 0       ,
\end{equation}
is studied in detail in Ref.~\cite{ref_Mazzolo_SAA}, where its asymptotic behavior is discussed.

\subsubsection{Conditioning with respect to the first-passage time density $\gamma^*(t)  =  \frac{a }{\sqrt{2 \pi t^3} } \frac{\cosh(\gamma a + \delta)}{\cosh(\delta)} e^{-\frac{a^2+\gamma ^2 t^2}{2 t}} $}
We now pursue the conditioning toward a prescribed first-passage-time density, taking as the target density that of a \emph{tanh-drift} process with new parameters $\gamma > 0$ and $\delta$, namely:
\begin{eqnarray}
\label{gammafirstpassagetanh-driftnew}
    \gamma^{*}_{[\gamma,\delta]}(t_2 \vert x_1,t_1) =  \frac{(a-x_1) }{\sqrt{2 \pi (t_2-t_1)^3} } \frac{\cosh(\gamma a + \delta)}{\cosh(\gamma x_1 + \delta)} e^{-\frac{(a-x_1)^2+\gamma ^2 (t_2-t_1)^2}{2 (t_2-t_1)}} \, .
\end{eqnarray}
Of course, for such a target density, all the relevant quantities have been computed previously, and since we already know that the survival probability is nonzero at infinity, we are dealing with the partial-survival case in the infinite-time limit (identical to the $\mu < 0$ case discussed in the preceding paragraph). Substituting Eqs.~(\ref{gammafirstpassagetanh-drift}), (\ref{prob_survival_tanh-drift}), and~(\ref{gammafirstpassagetanh-driftnew}) into Eq.~(\ref{Qinfinitypartial}), we obtain:
\begin{eqnarray}
 \label{Qinfinitypartialgammadelta}
    Q^{[\mathrm{partial}]}_{\infty [\gamma,\delta]}(x,t) & = &
    \int_t^{\infty} \gamma^*(T_a) \left[ \frac{\gamma(T_a \vert x,t)}{\gamma(T_a\vert 0,0)}  \right] dT_a  
    +  S^*(\infty ) \left[ \frac{S(\infty \vert x)}{S(\infty \vert 0)}  \right] \nonumber \\ 
                                                & = &
    \int_t^{\infty} \frac{a }{\sqrt{2 \pi T_a^3} } \frac{\cosh(\gamma a + \delta)}{\cosh(\delta)} e^{-\frac{a^2+\gamma ^2 T_a^2}{2 T_a}} \left[ \frac{T_a^{3/2}}{(T_a-t)^{3/2}} \frac{(a-x)}{a} \frac{\cosh (\beta )}{\cosh(\beta +\alpha  x)}    e^{ \frac{a^2+\alpha ^2 T_a^2}{2 T_a} }   e^{-\frac{(a-x)^2+\alpha ^2 (T_a - t)^2}{2 (T_a-t)} } \right] dT_a  \nonumber \\
                                                &  & + \left( 1 - e^{-a \gamma} \frac{\cosh(a \gamma + \delta)}{\cosh(\delta)} \right) \left[ \frac{\sinh (\alpha  (a-x))}{\sinh (a \alpha )}  \frac{ \cosh (\beta )}{\cosh (\alpha x + \beta)}  \right] \nonumber \\
                                                & = & 
    \frac{\cosh(\gamma a +\delta ) }{\cosh(\delta ) } \frac{\cosh(\beta )}{\cosh (\alpha x + \beta )}  e^{\gamma  (x-a)+\frac{1}{2} \left(\alpha ^2-\gamma ^2\right) t} +   \frac{\sinh (\alpha  (a-x))}{\sinh (a \alpha )}  \frac{ \cosh (\beta )}{\cosh (\alpha x + \beta)} \frac{ \sinh (a \gamma ) }{\cosh(\delta ) }  e^{-a \gamma -\delta } \nonumber \\
                                                & = &
    \frac{\cosh(\beta )}{\cosh (\alpha x + \beta )}  \frac{e^{-a \gamma -\delta }}{\cosh(\delta ) } \left(\cosh (a \gamma +\delta ) e^{\delta +\frac{1}{2}  (\alpha^2 -\gamma^2 )t+\gamma  x}+ \frac{\sinh (a \gamma )}{\sinh(a \alpha )} \sinh (\alpha  (a-x))\right)  \, ,
\end{eqnarray}
and the associated conditioned drift reads:
\begin{eqnarray}
\label{driftdoobgammadelta}
    \mu^*_{\infty [\gamma,\delta]}(x,t)  &=& \mu(x) +  \partial_x  \log Q^{[\mathrm{partial}]}_{\infty [\gamma,\delta]}(x,t) \nonumber \\
 & = &\alpha  \tanh (\alpha  x + \beta)  +  \partial_x \log \left[  \frac{\cosh(\beta )}{\cosh (\alpha x + \beta )}  \frac{e^{-a \gamma -\delta }}{\cosh(\delta ) } \left(\cosh (a \gamma +\delta ) e^{\delta +\frac{1}{2}  (\alpha^2 -\gamma^2 )t+\gamma  x}+ \frac{\sinh (a \gamma )}{\sinh(a \alpha )} \sinh (\alpha  (a-x))\right)  \right]
\nonumber \\
&=& \displaystyle{\frac{\gamma  \left(e^{2 (a \gamma +\delta )}+1\right) e^{\frac{1}{2} (\alpha^2 -\gamma^2 ) t+\gamma  x}-\frac{\alpha  \left(e^{2 a \gamma }-1\right) e^{\alpha  x} \left(e^{2 \alpha  (a-x)}+1\right)}{e^{2 a \alpha }-1}}{\left(e^{2 (a \gamma +\delta )}+1\right) e^{\frac{1}{2} (\alpha^2 -\gamma^2 ) t+\gamma  x}+\frac{\left(e^{2 a \gamma }-1\right) e^{\alpha  x} \left(e^{2 \alpha  (a-x)}-1\right)}{e^{2 a \alpha }-1}}  }\, .
\end{eqnarray}
Observe that this conditioned drift does not depend on the original parameter $\beta$. However, since it involves three parameters, $\alpha$, $\gamma$, and $\delta$, it cannot be associated with the previously considered drifts that depended on one or two parameters, except in the particular case $\gamma = \alpha$, where it reduces to:
\begin{eqnarray}
\label{driftdoobalphadelta}
    \mu^*_{\infty [\gamma,\delta]}(x,t)  = \alpha  \tanh (\alpha  x + \delta) ,
\end{eqnarray}
corresponding to a \emph{tanh-drift} process with parameters $\alpha$ and $\delta$. In the general case, the two processes—one with drift $\gamma \tanh(\gamma x + \delta)$ (corresponding to the \emph{tanh-drift} process) and the other with drift $\mu^*_{\infty[\gamma,\delta]}(x,t)$—share the same first-passage-time density to the barrier $a$, whatever the value of $a>0$. 
Consequently, based solely on the observation of first-passage times, one cannot determine whether the process originates from the \emph{tanh-drift} dynamics or from the new process introduced in this paragraph, a situation already encountered for the Brownian motion in the recent work of Ref.~\cite{ref_Mazzolo_SAA}.
For $\gamma \ne \alpha$ the stochastic representation of conditioned process, denoted as $X^*_{[\gamma,\delta]}(t)$, is not that of a \emph{tanh-drift} process with drift $\gamma \tanh(\gamma x + \delta)$ (as one might have expected) but obeys a complex inhomogeneous diffusion process:
\begin{align}
\label{SDEdriftdoobmuneglambda}
   dX^*_{[\gamma,\delta]}(t)  = \left( \frac{\gamma  \left(e^{2 (a \gamma +\delta )}+1\right) e^{\frac{1}{2} (\alpha^2 -\gamma^2 ) t+\gamma  X^*_{[\gamma,\delta]}(t)}-\frac{\alpha  \left(e^{2 a \gamma }-1\right) e^{\alpha  X^*_{[\gamma,\delta]}(t)} \left(e^{2 \alpha  (a-X^*_{[\gamma,\delta]}(t))}+1\right)}{e^{2 a \alpha }-1}}{\left(e^{2 (a \gamma +\delta )}+1\right) e^{\frac{1}{2} (\alpha^2 -\gamma^2 ) t+\gamma  X^*_{[\gamma,\delta]}(t)}+\frac{\left(e^{2 a \gamma }-1\right) e^{\alpha  X^*_{[\gamma,\delta]}(t)} \left(e^{2 \alpha  (a-X^*_{[\gamma,\delta]}(t))}-1\right)}{e^{2 a \alpha }-1}}  \right) dt  + dW(t) \, , ~~ t \geq 0  \,  ,
\end{align}
which retains a partial dependence on the original drift through the $\alpha$ term. Assuming that $X^*_{[\gamma,\delta]}(t)$ remains finite, the long-time behavior of this conditioned drift depends on the relative magnitudes of the two parameters, $\alpha$ and $\gamma$; more precisely:
\begin{eqnarray}
 \mu^*_{\infty [\gamma,\delta]}(x,t)   \underset{t \to \infty}{\sim\,} 
 \left\lbrace
  \begin{array}{lll}   
      -\alpha \coth(\alpha (a-x))
     &~~\mathrm{if~~} \gamma > \alpha 
    \\
       \, \, \, \, \gamma  
     &~~\mathrm{if~~} \gamma < \alpha 
  \end{array} \, .
\right.
\label{longtimegammadelta}
\end{eqnarray}
In the first case, $\gamma > \alpha$, when $x$ approaches the boundary $a$, the conditioned drift behaves as
\begin{equation}
\mu^*_{\infty [\gamma,\delta]}(x,\infty)  \underset{x \to a^-}{\sim}  -\frac{1}{a-x}
\end{equation}
which corresponds to the drift of the celebrated taboo process~\cite{ref_Knight,ref_Pinsky,ref_Mazzolo_Taboo}. Consequently, the conditioned process can never cross the barrier $a$, and can thus survive indefinitely. On the contrary, in the second case, when $\gamma < \alpha$, the behavior of the drift becomes counterintuitive. For large times, the conditioned drift $\gamma$ corresponds to a Brownian motion with a positive drift whose absorption at the boundary $a$ is certain, which contradicts the initial assumption of a nonzero survival probability. This suggests that, during intermediate times, some realizations of the process are pushed sufficiently far from the barrier $a$ to avoid absorption. A similar behavior has been reported for the type-II process described in Ref.~\cite{ref_Mazzolo_SAA}, which also corresponds to the process $X^*(t)$ discussed in the preceding paragraph. We can further characterize the process $X^*_{[\gamma,\delta]}(t)$ by computing its probability density. This quantity can be readily obtained, since the corresponding conditioned probability distribution is given by the product of the original propagator and the function $Q(x,t)$~\cite{ref_Monthus_Mazzolo,ref_deBruyne}, which, using Eqs.~(\ref{density_tanh-drift_abs}) and~(\ref{Qinfinitypartialgammadelta}), reads:
\begin{eqnarray}
\label{propagatorgammadelta}
    p_{\infty [\alpha,\delta]}^{*}(x,t) & = & Q^{[\mathrm{partial}]}_{\infty [\gamma,\delta]}(x,t) p_a(x,t) \nonumber \\
                                        & = &  \frac{\cosh(\beta )}{\cosh (\alpha x + \beta )}  \frac{e^{-a \gamma -\delta }}{\cosh(\delta ) } \left(\cosh (a \gamma +\delta ) e^{\delta +\frac{1}{2}  (\alpha^2 -\gamma^2 )t+\gamma  x}+ \frac{\sinh (a \gamma )}{\sinh(a \alpha )} \sinh (\alpha  (a-x))\right)  \nonumber \\
                                        &  & \times \frac{\cosh(\alpha x + \beta)}{\cosh(\beta)}  \frac{e^{- \frac{1}{2} \alpha^2 t}}{\sqrt{2 \pi t}} \left(e^{-\frac{x^2}{2t}} - e^{-\frac{(x - 2 a)^2}{2t}} \right)  \nonumber \\
                                        & = &  \frac{e^{-a\gamma}} {\cosh(\delta)}\,
                                        \left[
                                        e^{-\frac{\gamma^{2}t}{2} + \gamma x}\,\cosh(a\gamma + \delta)
                                        \;+\;
                                        e^{-\frac{\alpha^{2}t}{2} - \delta}\,\sinh(a\gamma)\,
                                        \frac{\sinh\!\big(\alpha(a - x)\big)}{\sinh(\alpha a)} 
                                        \right] \frac{1}{\sqrt{2\pi\,t}}\left(e^{-\frac{x^{2}}{2t}} - e^{-\frac{(x-2a)^{2}}{2t}}\right)  ,
\end{eqnarray}
an expression that is independent of the parameter $\beta$, as it should, since the drift $\mu^*_{\infty [\gamma,\delta]}(x,t)$ of the process $X^*_{[\gamma,\delta]}(t)$ is itself independent of $\beta$.
The propagator of the conditioned process has a rather complex expression that depends on the original parameter $\alpha$; however, it is worth noting that its survival probability at time $t>0$:
\begin{eqnarray}
\label{survivalgammadelta}
   S^*(t) & = & \int_{-\infty}^{a} p_{\infty [\alpha,\delta]}^{*}(x,t) \,dx \nonumber \\
                                        &=& \int_{-\infty}^{a} 
        \frac{e^{-a\gamma}} {\cosh(\delta)}
         \left[e^{-\frac{\gamma^{2}t}{2} + \gamma x} \cosh(a\gamma + \delta)+
          e^{-\frac{\alpha^{2}t}{2} - \delta} \sinh(a\gamma)\,
          \frac{\sinh\!\big(\alpha(a - x)\big)}{\sinh(\alpha a)} 
          \right] \frac{1}{\sqrt{2\pi\,t}}\left(e^{-\frac{x^{2}}{2t}} - e^{-\frac{(x-2a)^{2}}{2t}}\right)\,dx    \nonumber \\
                                        &=&  \,1- \frac{1}{2}\frac{\cosh(a\gamma+\delta)}{\cosh(\delta)}
                                            \left[e^{-a\gamma}\,\operatorname{erfc}\!\left(\frac{a-t\gamma}{\sqrt{2t}}\right)+ e^{a\gamma}\,\operatorname{erfc}\!\left(\frac{a+t\gamma}{\sqrt{2t}}\right)\right] \, ,
\end{eqnarray}
no longer depends on $\alpha$ and coincides with the survival probability of the \emph{tanh-drift} process $\gamma \tanh(\gamma x + \delta)$ [Eq.~(\ref{prob_survival_tanh-drift_up_to_T}) with initial position $x_1 = 0$ at time $t_1 = 0$], as expected.

\section{Conditioning the drifted Brownian motion}
\label{sec_Conditioned_BM}

In the same vein, we consider the conditioning of a Brownian motion with constant drift $\mu$, whose first-passage-time density is given by $\gamma_{\mu}(t) = \frac{a}{\sqrt{2\pi t^{3}}} e^{-\frac{(a - \mu t)^{2}}{2t}},\; t>0,$ first toward the first-passage-time density of the \emph{tanh-drift} process, and then toward the first-passage-time density of the taboo process.

Recall that the first-passage-time density of a Brownian motion with drift is normalized to unity when $\mu \ge 0$, and strictly less than one when $\mu < 0$ [Eq.(\ref{gammafirstBrowninte})]. Consequently, when conditioning, two distinct cases must be considered depending on the sign of $\mu$. Note, however, that since $S^*(\infty) = 1 - \frac{\cosh(\alpha a + \beta)}{\cosh(\beta)} e^{-a \alpha} < 1$ [Eq.~(\ref{prob_survival_tanh-drift})], the conditioning corresponds to partial survival in both cases.

\subsection{Conditioning the Brownian motion with drift $\mu$ with respect to the first-passage time density of the \emph{tanh-drift} process $\gamma^*(t)  =  \frac{a }{\sqrt{2 \pi t^3} } \frac{\cosh(\alpha a + \beta)}{\cosh(\beta)} e^{-\frac{a^2+\alpha ^2 t^2}{2 t}} $}
\subsubsection{Case $\mu \geq 0$}
When the initial drift is zero or positive, the function $Q_{\infty}^{[\mathrm{partial}]}(x,t)$, as well as its corresponding conditioned drift, have been derived in Ref.~\cite{ref_Monthus_Mazzolo}:
\begin{eqnarray}
Q^{[\mathrm{partial}]}_{\infty}(x,t)  =   \left( \frac{a-x}{a} \right)  e^{ - \mu x  + \frac{\mu^2 }{2} t} 
\left[ \int_t^{+\infty} \gamma^*(T_a)  \left(\frac{T_a}{T_a-t}  \right)^{\frac{3}{2} }   
e^{\frac{a^2}{2T_a} - \frac{(a-x)^2}{2(T_a-t)}} \, dT_a  
+ S^*(\infty )  \right]  \, ,
\label{Qmupos}
\end{eqnarray}
and
\begin{eqnarray}
 \mu^*_{\infty}(x,t) && = \mu +  \partial_x \log  Q^{[\mathrm{partial}]}_{\infty}(x,t) 
 \nonumber  \\  
&& = \frac{1}{x-a}  +  \partial_x \log  \left(  \int_t^{+\infty}  \gamma^*(T_a)  \left(\frac{T_a}{T_a-t}  \right)^{\frac{3}{2} }   
e^{\frac{a^2}{2T_a} - \frac{(a-x)^2}{2(T_a-t)}} dT_a 
+ S^*(\infty )  
\right) \, .
\label{driftdoobmupos}
\end{eqnarray}
Substituting the previously obtained expressions of $\gamma^*(T_a)$ and $S^*(\infty)$ into the expression for the drift, we obtain:
\begin{eqnarray}
 \mu^*_{\infty}(x,t) & = & \frac{1}{x-a}  +  \partial_x \log  \left(  \int_t^{+\infty} \frac{a}{\sqrt{2 \pi T_a^3}}\, \frac{\cosh(\alpha a + \beta)}{\cosh(\beta)}\, e^{-\frac{a^2 + \alpha^2 T_a^2}{2 T_a}}  \left(\frac{T_a}{T_a-t}  \right)^{\frac{3}{2} }   
e^{\frac{a^2}{2T_a} - \frac{(a-x)^2}{2(T_a-t)}} dT_a 
+ 1 - \frac{\cosh(\alpha a + \beta)}{\cosh(\beta)} e^{-a \alpha}  
\right)  \nonumber  \\  
& = & \frac{
a\alpha\,e^{\alpha x+\beta}\cosh(a\alpha+\beta)\;-\;\sinh(a\alpha)\,e^{\frac{\alpha^{2}t}{2}}
}{
a\,e^{\alpha x+\beta}\cosh(a\alpha+\beta)\;+\;\sinh(a\alpha)\,e^{\frac{\alpha^{2}t}{2}}(a-x)
} \, ,
\label{driftdoobmuposfinal}
\end{eqnarray}
which does not correspond to any previously established drift. However, the long-time behavior of this conditioned drift,
\begin{equation}
    \mu^*_{\infty}(x,t) \underset{t \to \infty}{\sim} -\frac{1}{a - x},
\end{equation}
coincides with the drift of the taboo process encountered earlier.

\subsubsection{Case $\mu < 0$}
\begin{figure}[h]
\centering
\includegraphics[width=3.5in,height=2.6in]{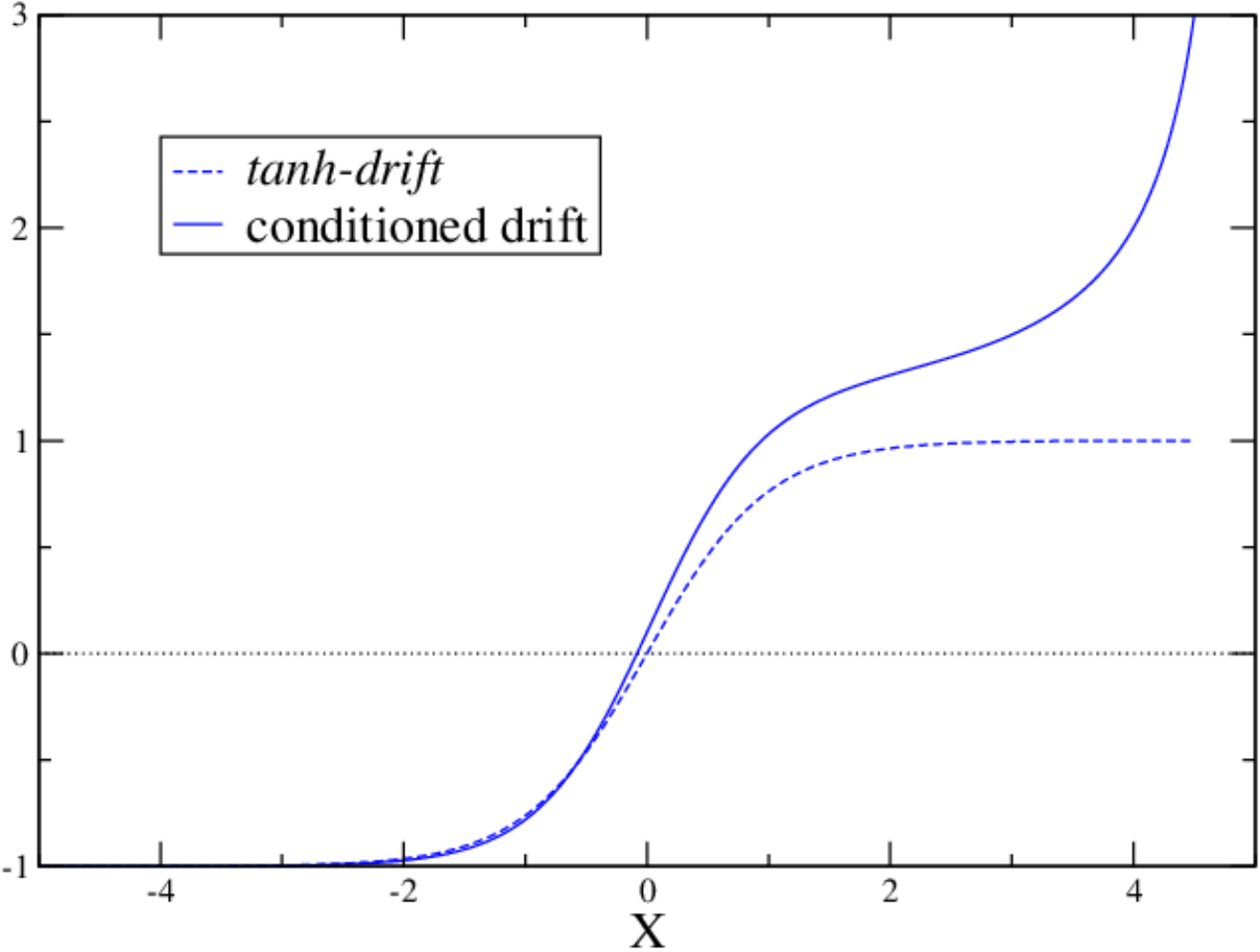}
\setlength{\abovecaptionskip}{15pt}  
\caption{ Plot of the conditioned drift with the parameter $\mu = -1$ and of the original \emph{tanh-drift} process with the parameters $\alpha = 1$ and $\beta = 0$, both with the barrier located at $a = 5$. The two drifts exhibit completely different behaviors for positive positions, which in practice could make it possible to distinguish the two processes.}
\label{fig:graph}
\end{figure}

When the initial drift is strictly negative, the function $Q_{\infty}^{[\mathrm{partial}]}(x,t)$ takes the following form~\cite{ref_Monthus_Mazzolo}:
\begin{eqnarray}
    Q_{\infty}^{[\mathrm{partial}]}(x,t) =  \left( \frac{a-x}{a} \right) e^{- \mu x + \frac{\mu^2}{2} t} 
\int_t^{+\infty} \gamma^*(T_a)  \left(\frac{T_a}{T_a-t}  \right)^{\frac{3}{2} }  
e^{\frac{a^2}{2T_a} - \frac{(a-x)^2}{2(T_a-t)}} dT_a  
+ S^*(\infty )  \left[ \frac{1- e^{2 \mu (a-x) } }{1- e^{2 \mu a } } \right] \, ,
\label{Qmuneg}
\end{eqnarray}
while the corresponding drift is still given by: 
\begin{eqnarray}
 \mu^*_{\infty}(x,t)  &=& \mu +  \partial_x \log  Q^{[\mathrm{partial}]}_{\infty}(x,t) 
 \nonumber  \\ 
& =  & 
\frac{
\int_t^{+\infty} \gamma^*(T_a)  \left(\frac{T_a}{T_a-t}  \right)^{\frac{3}{2} } \left( \frac{a-x}{a} \right)   
e^{- \mu x + \frac{\mu^2}{2} t +\frac{a^2}{2T_a} - \frac{(a-x)^2}{2(T_a-t)}}
\left[\frac{1}{x-a} +  \frac{a-x}{T_a-t}  \right] dT_a  
+ \mu \, S^*(\infty )  \left[ \frac{ 1+ e^{2 \mu (a-x) } }{1- e^{2 \mu a } } \right]
}
{
\int_t^{+\infty} \gamma^*(T_a)  \left(\frac{T_a}{T_a-t}  \right)^{\frac{3}{2} } \left( \frac{a-x}{a} \right)   
e^{- \mu x + \frac{\mu^2}{2} t +\frac{a^2}{2T_a} - \frac{(a-x)^2}{2(T_a-t)}} dT_a  
+ S^*(\infty )   \left[ \frac{1- e^{2 \mu (a-x) } }{1- e^{2 \mu a } } \right]
}
\label{driftdoobmuneg}
\end{eqnarray}
Substituting once again the previously obtained expressions of $\gamma^*(T_a)$ and $S^*(\infty)$ into the expression for the drift, we get:
\begin{eqnarray}
 \mu^*_{\infty}(x,t) = \frac{\frac{2a\,e^{a\alpha+\beta}\cosh(a\alpha+\beta)\,
e^{-\tfrac{1}{2}(\alpha^2-\mu^2)t}\,e^{\alpha x}}{a-x}
\left(\alpha+\frac{1}{a-x}\right)
\;-\;
2\mu\,e^{a\alpha}\,\frac{\sinh(a\alpha)}{\sinh(a\mu)}\,
\cosh\!\bigl(\mu(a-x)\bigr)}
{\frac{2a\,e^{a\alpha+\beta}\cosh(a\alpha+\beta)\,
e^{-\tfrac{1}{2}(\alpha^2-\mu^2)t}\,e^{\alpha x}}{a-x}
\;+\;
2e^{a\alpha}\,\frac{\sinh(a\alpha)}{\sinh(a\mu)}\,
\sinh\!\bigl(\mu(a-x)\bigr)}  \, ,
\end{eqnarray}
an expression that depends on the three parameters $\mu$, $\alpha$, and $\beta$. The asymptotic long-time behavior of the drift depends on the relative magnitude of the parameters $\mu$ and $\alpha$, more precisely:
\begin{eqnarray}
\mu^*_{\infty}(x,t) \underset{t \to \infty}{\sim}
\begin{cases}
-\mu\,\coth\!\bigl(\mu(a-x)\bigr), & |\mu|<\alpha,\\[6pt]
\alpha+\dfrac{1}{a-x}, & |\mu|>\alpha,\\[8pt]
\end{cases}
\end{eqnarray}
For $\alpha = -\mu$, the expression of the conditioned drift becomes time-independent and reads:
\begin{eqnarray}
\mu^*_{\infty}(x) =
\frac{\displaystyle
\frac{2a\,e^{a|\mu|+\beta}\cosh(a|\mu|+\beta)\,e^{|\mu| x}}{a-x}\!\left(|\mu|+\frac{1}{a-x}\right)
\;-\;2\alpha\,e^{a|\mu|}\cosh\!\bigl(|\mu|(a-x)\bigr)}
{\displaystyle
\frac{2a\,e^{a|\mu|+\beta}\cosh(a|\mu|+\beta)\,e^{|\mu| x}}{a-x}
\;+\;2e^{a|\mu|}\,\sinh\!\bigl(|\mu|(a-x)\bigr)} \, ,
\end{eqnarray}
which simplifies further when choosing $\beta = 0$:

\begin{eqnarray}
   \mu^*_{\infty}(x) =         \frac{a + a \mu (x-a) + \mu (a-x)^{2} e^{2\mu x} + (a + \mu x^2 - a\mu x) e^{2 a\mu} }{ (a-x)\Big(a + (a-x)\,e^{2\mu x}+ x\,e^{2 a\mu} \Big)}\, .
\end{eqnarray}
Figure~\ref{fig:graph} illustrates the behavior of this drift together with the corresponding \emph{tanh-drift}.

\subsection{Conditioning the Brownian motion with drift $\mu$ with respect to the first-passage time density of the taboo process $\gamma^*(t) = (b-a)\frac{a}{b} \frac{1}{\sqrt{2 \pi t^3}} e^{-\frac{a^2}{2 t}} $}

Analytical expressions for the fundamental quantities of the taboo process, such as the first-passage-time density and the survival probability, will be derived in the next section, where it is shown that [Eq.\eqref{gammafirstpassagetaboo}]:
\begin{equation}
\label{gammastarfirstpassagetabooinfini}
   \gamma(T_a) = (b-a)\frac{a}{b} \frac{1}{\sqrt{2 \pi T_a^3}} e^{-\frac{a^2}{2 T_a}}  \, ,
\end{equation}
and [Eq.\eqref{prob_survival_taboo_up_to_infinity}]
\begin{eqnarray}
\label{Sstar_taboo_up_to_infinity}
   S(\infty)  = \frac{a}{b} \, .
\end{eqnarray}

\subsubsection{Case $\mu \geq 0$}
Substituting the expressions of the first-passage-time density and the survival probability of the taboo process into Eq.~\eqref{Qmupos} yields
\begin{eqnarray}
Q^{[\mathrm{partial}]}_{\infty}(x,t) & = &  \left( \frac{a-x}{a} \right)  e^{ - \mu x  + \frac{\mu^2 }{2} t} 
\left[ \int_t^{+\infty}    (b-a)\frac{a}{b} \frac{1}{\sqrt{2 \pi T_a^3}} e^{-\frac{a^2}{2 T_a}}  \left(\frac{T_a}{T_a-t}  \right)^{\frac{3}{2} }   
e^{\frac{a^2}{2T_a} - \frac{(a-x)^2}{2(T_a-t)}} \, dT_a + \frac{a}{b} \right]  \nonumber \\
                            & = & \frac{(b-x)}{b} e^{ - \mu x  + \frac{\mu^2 }{2} t} \, ,
\label{Qmupostotaboo}
\end{eqnarray}
and
\begin{eqnarray}
 \mu^*_{\infty}(x,t) = \mu +  \partial_x \log  Q^{[\mathrm{partial}]}_{\infty}(x,t) = - \frac{1}{b-x} \, 
\label{driftdoobmupostotaboo}
\end{eqnarray}
which is the drift of the taboo process with taboo state $b$.

\subsubsection{Case $\mu < 0$}
Likewise, substituting the expressions of the first-passage-time density and the survival probability of the taboo process into Eq.~\eqref{Qmuneg} leads to
\begin{eqnarray}
    Q_{\infty}^{[\mathrm{partial}]}(x,t) & = &  \left( \frac{a-x}{a} \right) e^{- \mu x + \frac{\mu^2}{2} t} 
\int_t^{+\infty}  (b-a)\frac{a}{b} \frac{1}{\sqrt{2 \pi T_a^3}} e^{-\frac{a^2}{2 T_a}}  \left(\frac{T_a}{T_a-t}  \right)^{\frac{3}{2} }  
e^{\frac{a^2}{2T_a} - \frac{(a-x)^2}{2(T_a-t)}} dT_a  
+ \frac{a}{b}  \left[ \frac{1- e^{2 \mu (a-x) } }{1- e^{2 \mu a } } \right] \nonumber \\
                                        & = & \frac{(b-a)}{b}  e^{- \mu x + \frac{\mu^2}{2} t} + \frac{a}{b}  \left( \frac{1- e^{2 \mu (a-x) } }{1- e^{2 \mu a } } \right)
\label{Qmunegtotaboo}
\end{eqnarray}
and for the corresponding drift 
\begin{eqnarray}
 \mu^*_{\infty}(x,t) & = & \mu +  \partial_x \log  Q^{[\mathrm{partial}]}_{\infty}(x,t) 
 \nonumber  \\ 
                     & = & \mu + \frac{-\mu  (b-a) e^{\frac{\mu ^2 t}{2}-\mu  x}+\frac{2 a \mu  e^{2 \mu  (a-x)}}{\left(1-e^{2 a \mu }\right)}}{{(b-a) e^{\frac{\mu ^2 t}{2}-\mu  x}}+\frac{a \left(1-e^{2 \mu  (a-x)}\right)}{ \left(1-e^{2 a \mu }\right)}} \, .
\label{driftstarmunegtotaboo}
\end{eqnarray}
Observe that when taboo state $b$ approaches the boundary $a$, the conditioned drift behaves as
\begin{equation}
      \mu^*_{\infty}(x,t)  \underset{b \to a^+}{\sim}  -\mu  \coth (\mu  (a-x)) \, ,
\end{equation}
and when $\mu \to 0^-$ 
\begin{equation}
      \mu^*_{\infty}(x,t)  \underset{\mu \to 0^-}{\sim} -\frac{1}{b-x} \, ,
\end{equation}
as expected, thereby recovering the previously obtained case $\mu = 0$.

\section{Conditioning the taboo process}
\label{sec_Conditioned_taboo}
Having established that several conditioned \emph{tanh-drift} processes develop, near
the absorbing boundary, a drift identical to that of the taboo diffusion, it is
natural to investigate the conditioning of the taboo process itself. Our goal in
this section is therefore twofold: first, to derive explicit expressions for the
propagator, the first-passage-time density, and the survival probability of the
taboo process using Girsanov’s theorem; and second, to examine how these
quantities transform under various conditioning schemes, namely conditioning on
a finite time horizon, conditioning on eternal survival, and conditioning on the
first-passage-time density of a Brownian motion with constant drift. The
resulting picture further strengthens the structural parallels between taboo
dynamics, the \emph{tanh-drift} process, and the drifted Brownian motion.
To proceed, we start from a standard Brownian motion with an absorbing barrier at $x = a>0$, whose propagator reads [Eq.~\eqref{density_BM_abs}]:
\begin{equation}
	 p_{\mathrm{a}}(x_2,t_2\vert x_1,t_1)  =  \frac{1}{\sqrt{2 \pi (t_2 - t_1)}} \left\{e^{-\frac{(x_2 - x_1)^2}{2(t_2 - t_1)}} - e^{-\frac{(x_1 + x_2 - 2 a)^2}{2(t_2 - t_1)}} \right\},~t_2>t_1>0 .
\end{equation}
We will then derive the corresponding quantity for the taboo process in the semi-infinite state space $(-\infty,b)$ (with a taboo state $b>a$), by imposing, in space, the drift of a taboo process, namely
\begin{equation}
    \mu(x) = -\frac{1}{b - x} \, .
\end{equation}
For such a drift, the weighting factor $Z(t)$ [Eq.~\eqref{def_Z}] reads:  
\begin{equation}  
\label{Z_taboo_def}  
	 Z(t) = e^{\textstyle \int_{t_1}^{t_2} -\frac{dW(u)}{b - W(u)}
      - \frac{1}{2} \int_{t_1}^{t_2} \frac{du}{(b - W(u))^2}} .  
\end{equation}
Applying Itô's lemma to the function $\log(b - W(t))$, namely
\begin{equation} 
    d\log(b - W(t))
    = -\,\frac{dW(t)}{b - W(t)}
      - \frac{dt}{2\, (b - W(t))^2} ,
\end{equation}
allows us to write the weighting factor in the simpler form
\begin{equation}  
\label{Z_taboo_final}  
	 Z(t)
   =  e^{{\big{[}} \textstyle \log\!\big(b-W(u)\big) {\big{]}}_{t_1}^{t_2}}
   = \frac{b - W(t_2)}{b - W(t_1)} \,  ,
 \end{equation}
and the transition density of the taboo process (with taboo state $b$) absorbed at the level $a$ can be obtained in closed form and reads:
\begin{equation}
\label{proptaboowithabsorption}
	 p_{\mathrm{a}}^{taboo}(x_2,t_2\vert x_1,t_1)  =   \frac{(b - x_2)}{(b - x_1)} \times \frac{1}{\sqrt{2 \pi (t_2 - t_1)}} \left\{e^{-\frac{(x_2 - x_1)^2}{2(t_2 - t_1)}} - e^{-\frac{(x_1 + x_2 - 2 a)^2}{2(t_2 - t_1)}} \right\},~t_2>t_1>0 .
\end{equation}
Equipped with this result, it is then straightforward to determine the first-passage-time density $\gamma(t_2 \mid x_1,t_1)$ of the process:
\begin{equation}
\label{gammafirstpassagetaboo}
   \gamma(t_2 \vert x_1,t_1) = - \frac{1 }{2}  \left( \partial_{x_2} p_{\mathrm{a}}^{taboo}(x_2,t_2 \vert x_1,t_1) \right)\bigg\vert_{x_2=a} = (b-a)\frac{(a- x_1)}{(b - x_1)} \frac{1}{\sqrt{2 \pi (t_2 - t_1)^3}} e^{-\frac{(a - x_1)^2}{2(t_2 - t_1)}}  .
\end{equation}
The survival probability of the process up to time $T$ follows:
\begin{eqnarray}
\label{prob_survival_taboo_up_to_T}
   S(T \vert x_1,t_1)  = 1 -  \int_{t_1}^{T} \gamma(t_2 \vert x_1,t_1) \, dt_2 =\frac{a-x_1 +(b-a) \text{erf}\left(\frac{a-x_1}{\sqrt{2(T-t_1)}}\right)}{b-x_1} ,
\end{eqnarray}
from which we immediately obtain:
\begin{eqnarray}
\label{prob_survival_taboo_up_to_infinity}
   S(\infty \vert x_1)  = \frac{a-x_1}{b-x_1} .
\end{eqnarray}
Thus, the forever-survival probability takes a simple form. Observe that as $b$ approaches $a$, the drift of the taboo process increasingly repels the trajectory from the barrier, and the process can no longer come close to it. In this case, the survival becomes certain, and $\lim_{b \to a} S(\infty \vert x_1,t_1) = 1$, as expected.
\noindent With these quantities at hand, we now have everything required to condition the taboo process to have a prescribed first-passage-time density.

\subsection{Conditioning on a finite time horizon $T$}
The conditioning on a finite time horizon $T$ was established in Section~\ref{subsec_conditoning_finite_horizon}. 
Recall that the conditioned drift is given by
\begin{equation}
    \mu^*_T(x,t)
    = \mu(x) + \partial_x \log Q_T(x,t)
    = -\frac{1}{\,b - x\,} + \partial_x \log Q_T(x,t),
\end{equation}
where $Q_T(x,t)$ is defined in Eq.~\eqref{QsiniteT}:
\begin{equation}
    Q_T(x,t)
    =
    \int_t^{T} \gamma^*(T_a)\,
        \frac{\gamma(T_a \mid x,t)}{\gamma(T_a \mid 0,0)} dT_a\ 
    +
    \int_{-\infty}^{a}  P^*(y,T)\,
        \frac{ p_a(y,T \mid x,t) }{ p_a(y,T \mid 0,0) } dy.
\end{equation}
Substituting the explicit expressions of the first-passage density 
$\gamma(t_2 \mid x_1,t_1)$ [Eq.~\eqref{gammafirstpassagetaboo}] 
and of the absorbed taboo propagator 
$p_{\mathrm{a}}^{taboo}(x_2,t_2\vert x_1,t_1)$ [Eq.~\eqref{proptaboowithabsorption}] 
into the formula above, we obtain:
\begin{eqnarray}
 \label{QsiniteTsimplifiedtaboo}
    Q_T(x,t) = \frac{b}{(b - x)} \bigg[ \displaystyle \frac{(a-x)}{a} \int_t^{T} \gamma^*(T_a) \sqrt{\frac{T_a^3}{(T_a-t)^3}} e^{-\frac{(a-x)^{2}}{2(T_a-t)} + \frac{a^{2}}{2 T_a}} dT_a   
 + 
    \sqrt{\frac{T}{T-t}} \int_{-\infty}^a P^*(y,T )  \frac{
\left(e^{-\frac{(x-y)^{2}}{2 (T-t)}}-e^{-\frac{(2 a-x-y)^{2}}{2 (T-t)}}\right)}
{\left(e^{-\frac{y^{2}}{2 T}}-e^{-\frac{(y-2 a)^{2}}{2 T}}\right)} dy \bigg] \, .
\end{eqnarray}
And for the conditioned drift:
\begin{eqnarray}
\label{driftfiniteTsimplifiedtaboo}
    \mu^*_T(x,t)  = \partial_x \log \left[ \displaystyle  \frac{(a-x)}{a}  \int_t^{T} \gamma^*(T_a) \sqrt{\frac{T_a^3}{(T_a-t)^3}} e^{-\frac{(a-x)^{2}}{2(T_a-t)} + \frac{a^{2}}{2 T_a}} dT_a   \displaystyle  + 
    \sqrt{\frac{T}{T-t}}  \int_{-\infty}^a P^*(y,T )  \frac{
\left(e^{-\frac{(x-y)^{2}}{2 (T-t)}}-e^{-\frac{(2 a-x-y)^{2}}{2 (T-t)}}\right)}
{\left(e^{-\frac{y^{2}}{2 T}}-e^{-\frac{(y-2 a)^{2}}{2 T}}\right)} dy  \right] \, ,
\end{eqnarray}
which is precisely the expression obtained when conditioning the \emph{tanh-drift} process over a finite time horizon [Eq.~\eqref{driftfiniteTsimplified}]. 
Consequently, all the results derived in Section~\ref{subsec_conditoning_finite_horizon} directly carry over to the present case, in particular when the process is conditioned to be fully absorbed at a single prescribed time $\gamma^*(T_a) = \delta(T_a - T^*)$, the conditioned drift is given by [Eq.~\eqref{driftdoobdeltaT}]:
\begin{eqnarray}
\mu^*(x,t) = -\frac{1}{a-x} +  \frac{a-x}{T^*-t} \, .
\end{eqnarray}
Note that, as in Section~\ref{subsec_conditoning_finite_horizon}, when $a$ becomes large the conditioned drift reduces to that of a Brownian bridge,
\begin{equation}
    \mu^{[\mathrm{BB}]}(x,t) = \frac{a - x}{T^* - t},
\end{equation}
which might seem to contradict the result of Benjamini and Lee~\cite{ref_Benjamini}, who proved that only Brownian motion with constant drift and the \emph{tanh-drift} process share the same bridge. However, a closer examination of their work shows that their result holds under the assumption that the drifts do not blow up. More precisely, drifts sharing the same bridges satisfy the equation $\mu(x) + \mu(x)^2 = k$, and the authors retained only the solution with $k > 0$, corresponding to non-exploding drifts. If this condition is relaxed, then the taboo drift $\mu(x) = -1/(b - x)$ also satisfies the equation with $k = 0$, and there is no contradiction.

\subsection{Conditioning on full survival as $t \to \infty$}
\label{sec_Conditioned_taboo_full_survival}
The conditioning toward a forever-surviving process was established in Section~\ref{sec_Conditioned_tanh-drift_full_survival}, where the conditioned drift is given by
\begin{equation}
    \mu^*_T(x,t)
    = \mu(x) + \partial_x \log Q_{\infty}^{[\mathrm{surviving}]}(x,t)
    = -\frac{1}{b - x} + \partial_x \log Q_{\infty}^{[\mathrm{surviving}]}(x,t),
\end{equation}
with $Q_{\infty}^{[\mathrm{surviving}]}(x,t)$ defined in Eq.~\eqref{Qspacesurvivalinfinity_def} as
\begin{equation}
    Q_{\infty}^{[\mathrm{surviving}]}(x,t)
    = \frac{S(\infty \vert x)}{S(\infty \vert 0)} \, .
\end{equation}
Substituting the simple expression of the forever-survival probability $S(\infty \vert \cdot)$ for the taboo process [Eq.~\eqref{prob_survival_taboo_up_to_infinity}], we immediately obtain
\begin{equation}
    \mu^*_T(x,t) = -\frac{1}{a - x} \, .
\end{equation}
Therefore, conditioning a taboo process living in $(-\infty,b)$ to survive forever in a smaller region $(-\infty,a)$ (with $a < b$) yields another taboo process with taboo state $a$.

\subsection{Conditioning the taboo process on the first-passage-time density of Brownian motion with constant drift}
Finally, we focus on conditioning the taboo process on the first-passage-time density of a Brownian motion with constant drift $\mu$ through the barrier $a$,
\begin{equation}
\label{gammafirstpassageBMwithdrift}
     \gamma^*(t) = \frac{a}{\sqrt{2\pi t^3}}\, e^{-\frac{(a - \mu t)^2}{2t}} \, ,
\end{equation}
distinguishing, as required, between the two cases $\mu \ge 0$ and $\mu < 0$.

\bigskip
\paragraph*{Case $\mu \geq 0$: Vanishing survival in the infinite-time limit}\mbox{}\\
When the conditioning is toward a first-passage-time distribution $\gamma^*(T_a)$ that is normalized to unity, the function $Q_{\infty}^{[\mathrm{time}]}(x,t)$ is given by Eq.~\eqref{Qtimeinfinity}. Substituting the expressions of $\gamma(t_2 \vert x_1,t_1)$ [Eq.~\eqref{gammafirstpassagetaboo}] and $\gamma^*(t)$ [Eq.~\eqref{gammafirstpassageBMwithdrift}] into this formula yields
\begin{eqnarray}
    Q_{\infty}^{[time]}(x,t) = \int_t^{+\infty} \frac{a}{\sqrt{2\pi T_a^3}}\, e^{-\frac{(a - \mu T_a)^2}{2 T_a}} \frac{(b-a)\frac{(a- x)}{(b - x)} \frac{1}{\sqrt{2 \pi (T_a - t)^3}} e^{-\frac{(a - x)^2}{2(T_a - t)}}}{(b-a)\frac{a}{b} \frac{1}{\sqrt{2 \pi T_a^3}} e^{-\frac{a^2}{2 T_a}}} dT_a = \frac{b \, e^{\mu  x-\frac{\mu ^2 t}{2}}}{b-x} \, .
 \label{QtimeinfinitytabootoBM}
\end{eqnarray}
The corresponding conditioned drift is thus given by:
\begin{eqnarray}
\mu^*_{\infty}(x,t)  = \mu(x) +  \partial_x  \log Q_{\infty}^{[time]}(x,t) = -\frac{1}{b - x} +  \partial_x  \log \left( \frac{b \, e^{\mu  x-\frac{\mu ^2 t}{2}}}{b-x} \right) = \mu  \, .
\label{driftdoobforevertabootoBMmupositive}
\end{eqnarray}
Therefore, conditioning the taboo process on the normalized first-passage-time distribution of a Brownian motion with a uniform positive drift $\mu$ results in a conditioned process that is itself a Brownian motion with drift $\mu$, exactly as in the case of the initial \emph{tanh-drift} process.

\medskip
\paragraph*{Case $\mu < 0$: Partial survival in the infinite-time limit}\mbox{}\\
When the conditioning is performed with respect to a first-passage-time distribution $\gamma^*(T_a)$ that is not normalized to unity, the function $Q_{\infty}^{[\mathrm{partial}]}(x,t)$ is given by Eq.~\eqref{Qinfinitypartial}:
\begin{eqnarray}
    Q_{\infty}^{[\mathrm{partial}]}(x,t)
    = \int_t^{\infty} \gamma^*(T_a)\,
      \frac{\gamma(T_a \vert x,t)}{\gamma(T_a\vert 0,0)}\, dT_a
      + S^*(\infty)\, \frac{S(\infty \vert x)}{S(\infty \vert 0)} \, .
\end{eqnarray}
Inserting the expressions of $\gamma(t_2 \vert x_1,t_1)$ [Eq.~\eqref{gammafirstpassagetaboo}], 
$\gamma^*(t)$ [Eq.~\eqref{gammafirstpassageBMwithdrift}], 
$S(\infty \vert x_1)$ [Eq.~\eqref{prob_survival_taboo_up_to_infinity}], 
and $S^*(\infty)$ [Eq.~\eqref{survivalBrown}] 
into this formula leads to:
\begin{eqnarray}
    Q_{\infty}^{[\mathrm{partial}]}(x,t)
    &=& \int_t^{+\infty} \frac{a}{\sqrt{2\pi T_a^3}}\, e^{-\frac{(a - \mu T_a)^2}{2 T_a}} \frac{(b-a)\frac{(a- x)}{(b - x)} \frac{1}{\sqrt{2 \pi (T_a - t)^3}} e^{-\frac{(a - x)^2}{2(T_a - t)}}}{(b-a)\frac{a}{b} \frac{1}{\sqrt{2 \pi T_a^3}} e^{-\frac{a^2}{2 T_a}}} dT_a + \left( 1- e^{2 \mu (a-x)} \right) \frac{\left(\frac{a-x}{b-x}\right)}{\left(\frac{a}{b}\right)} \nonumber \\
    &=& \frac{b e^{-\frac{1}{2} \mu  (-4 a+\mu  t+2 x)}}{b-x}+\frac{b \left(1-e^{2 a \mu }\right) (a-x)}{a (b-x)} \, ,
\end{eqnarray}
and the conditioned drift is expressed as:
\begin{eqnarray}
\mu^*_{\infty}(x,t)  =\mu(x) +  \partial_x  \log Q_{\infty}^{[time]}(x,t) 
                     &=&-\frac{1}{b - x} +  \partial_x  \log \left( \frac{b e^{-\frac{1}{2} \mu  (-4 a+\mu  t+2 x)}}{b-x}+\frac{b \left(1-e^{2 a \mu }\right) (a-x)}{a (b-x)} \right) \nonumber \\
                     &=& \frac{2\sinh(a\mu)\,e^{\frac{\mu^2 t}{2}+\mu(x-a)} - a\mu}{a - 2(a-x)\sinh(a\mu)\,e^{\frac{\mu^2 t}{2}+\mu(x-a)}}  \, ,
\label{driftdoobforevertabootoBMmunegative}
\end{eqnarray}
yielding a drift that differs from all previously derived conditioned drifts, and thus corresponds to a genuinely new process, governed by an inhomogeneous diffusion:
\begin{equation}
\label{eq_diffusion_conditioned_tabootoBMmunegative}
	  dX^*(t) = \left[ \frac{2\sinh(a\mu)\,e^{\frac{\mu^2 t}{2}+\mu(X^*(t)-a)} - a\mu}{a - 2(a-X^*(t))\sinh(a\mu)\,e^{\frac{\mu^2 t}{2}+\mu(X^*(t)-a)}}  \right] dt + dW(t), \qquad t \ge 0 ,
\end{equation}
which no longer depends on the taboo state $b$. Observe that the long-time behavior of this conditioned drift,
\begin{equation}
    \mu^*_{\infty}(x,t) \underset{t \to \infty}{\sim} -\frac{1}{a - x},
\end{equation}
coincides with the drift of the taboo process, with taboo state $a$.

\section{Conclusion}
\label{sec_Conclusion} 
By conditioning the \emph{tanh-drift} process on various first-passage-time distributions, we obtained exact expressions for the associated conditioned drifts, and therefore for the associated stochastic differential equations, which in turn enable efficient numerical simulations.
The results depend strongly on whether the conditioning is imposed at a finite time or at an infinite time. In the latter case, the conditioned processes do not coincide with the original \emph{tanh-drift} process, thereby extending to this setting a phenomenon already observed for Brownian motion: namely, the existence of processes that differ from the unconditioned one (here the \emph{tanh-drift} process) while sharing exactly the same first-passage-time density.

In contrast, the situation is radically different when conditioning at a finite time, for which we found that the conditioned \emph{tanh-drift} process exhibits exactly the same behavior as Brownian motion motion with drift under the same conditioning. This result generalizes an elegant theorem of Benjamini and Lee, who proved that Brownian motion motion with drift and the \emph{tanh-drift} process share the same Brownian bridge—and, as shown here, also the same constrained bridge conditioned to remain below a given threshold.

Beyond the \emph{tanh-drift} diffusion itself, our analysis revealed that several conditioned  Bene\v{s} drifts converge near the absorbing boundary to the drift of the taboo diffusion. This observation motivated a parallel study of the taboo process, for which we used Girsanov’s theorem to derive the propagator, the first-passage-time distribution, the survival probability, and various conditioned versions. In particular, we showed that conditioning a taboo process to survive forever in a smaller domain yields again a taboo process with a shifted taboo state, while conditioning it --- in the case of partial survival in the infinite-time limit --- on the first-passage-time law of a drifted Brownian motion produces a genuinely new inhomogeneous diffusion whose long-time drift converges to the taboo drift.

Another natural question raised by the present work is whether the conditioning procedure can also be used in reverse, namely to determine whether a given complicated drift may be interpreted as the conditioned version of a simpler reference diffusion. In general, this amounts to a non-trivial inverse problem for the conditioning information function and cannot be expected to hold universally. However, we believe that it may be possible for certain structured classes of diffusions, and that identifying such classes would constitute an interesting direction for future work.

We would like to conclude our study by emphasizing the importance of Girsanov’s theorem. Not only does it allow one to derive exact expressions for the probability densities of complex diffusion processes -as we obtained for the \emph{tanh-drift} process with absorbing conditions- but it also preserves the underlying structural origin of the process.  
This is precisely the difference between the expression of the first-passage-time density of Brownian motion with drift given in Eq.~\eqref{density_BM_abs_Girsanov} and the standard textbook formula in Eq.~\eqref{density_BM_abs_images}, in which the underlying structure inherited from standard Brownian motion is obscured. We believe that systematically employing this theorem will help to reveal deep structural connections between seemingly different stochastic processes.



\newpage
\appendix

\section{Summary Tables of Conditioning Results}
\label{appendix_tables}
In this appendix, we present two tables summarizing our main conditioning results. 
The first table collects the results corresponding to conditioning toward absorption at a fixed time $T^*$ at the level $a$, as well as conditioning toward full survival. 
The second table compiles the results obtained when the conditioning is performed with respect to a prescribed first-passage-time (FPT) density, whether normalized or not. 
In each case, for a given original drift, we specify the imposed FPT density together with the resulting drift of the conditioned process. 

In the following, we denote by $\mathrm{BM}(\mu)$ a Brownian motion with constant drift $\mu$, by $\tanh(\alpha, \beta)$ the \emph{tanh-drift} process, i.e. the process with drift $\mu(x) = \alpha \tanh(\alpha x + \beta)$ and by $\mathrm{taboo(a)}$ the taboo process with taboo state $a$, 
i.e. the process with drift $\mu(x) = -1/(a-x)$.
\begin{table}[!ht]
  \caption{
    Drift of the original process, conditioning on absorption at $a$ at a fixed time $T^*$ or on forever survival, and the induced drift of the conditioned process.
  }
  \label{table_fullsurvival}
  \begin{ruledtabular}
    \begin{tabular}{ccc}
      \parbox[t]{0.25\textwidth}{
        original drift 
      }
      & target first-passage-time density
      & drift of the conditioned process \\
      \hline

      \noalign{\vskip 4pt} 
      $\mathrm{BM}(\mu)$
      &
      $\delta(T_a - T^*)$
      &
      $\displaystyle
      -\frac{1}{a-x} +  \frac{a-x}{T^*-t} 
      $
      \\
      \noalign{\vskip 8pt} 
      $\displaystyle
      \mathrm{BM}(\mu \ge 0)
      $
      &
      forever survival
      &
      $\displaystyle
      \mathrm{taboo}(a)$
      \\
      \noalign{\vskip 8pt} 
      $\mathrm{BM}(\mu < 0) $
      &
      forever survival
      &
      $\displaystyle
       -\mu \coth(\mu (a-x)) 
      $
      \\
      \noalign{\vskip 8pt}                   
      \noalign{\vskip 4pt} 
      $\tanh(\alpha, \beta)$
      &
      $\delta(T_a - T^*)$
      &
      $\displaystyle
      -\frac{1}{a-x} +  \frac{a-x}{T^*-t} 
      $
    
      \\
      \noalign{\vskip 8pt} 
      $\tanh(\alpha, \beta)$
      &
      forever survival
      &
      $\displaystyle
       -\alpha \coth(\alpha (a-x)) 
      $
      \\
      \noalign{\vskip 8pt} 
      $\displaystyle
      \mathrm{taboo}(b)
      $
      &
      $\delta(T_a - T^*)$
      &
      $\displaystyle
      -\frac{1}{a-x} +  \frac{a-x}{T^*-t} 
      $
      \\
      \noalign{\vskip 8pt} 
      $\displaystyle
      \mathrm{taboo}(b)
      $
      &
      forever survival
      &
      $\mathrm{taboo}(a)$
      \\
    \end{tabular}
  \end{ruledtabular}
\end{table}
\FloatBarrier
Table~\ref{table_fullsurvival} summarizes the conditioned drifts obtained under the two conditioning schemes of absorption at a fixed time $T^*$ and full survival. 
The first three rows of this table correspond to Brownian motion with constant drift; they are taken from Ref.~\cite{ref_Monthus_Mazzolo} and are recalled here for completeness.

Beyond collecting known results, this table highlights remarkable structural relations between the conditioned drifts of three different processes: Brownian motion with drift, the \emph{tanh-drift} (Bene\v{s}) process, and the taboo process. In particular, when conditioning on absorption at a fixed time $T^*$, all three processes lead to the same drift,
\begin{equation}
-\frac{1}{a-x} + \frac{a-x}{T^*-t} \, ,
\end{equation}    
which corresponds to that of a Brownian bridge conditioned to reach the level $a$ at time $T^*$ while remaining under $a$ at earlier times.

In contrast, when conditioning on full survival, the resulting conditioned drifts depend on the original dynamics but retain a common repulsive structure near the absorbing boundary. Specifically, the conditioned drift reduces to the taboo drift $-1/(a-x)$ for the taboo process and for Brownian motion with nonnegative drift, while it takes similar hyperbolic forms for Brownian motion with negative drift and for the \emph{tanh-drift} process, respectively. These results emphasize deep connections between these conditioned processes and illustrate how different unconditioned dynamics may lead to closely related, or even identical, conditioned behaviors.

\begin{table}[!ht]
  \caption{
    Drift of the original process, imposed FPT density, and the induced drift of the conditioned process.
  }
  \label{table_densities}
  \begin{ruledtabular}
    \begin{tabular}{ccc}
      \parbox[t]{0.25\textwidth}{
        original drift 
      }
      & target first-passage-time density
      & drift of the conditioned process \\
      \hline

      \noalign{\vskip 8pt} 
      $\tanh(\alpha, \beta)$ 
      &
      $\displaystyle
      \mathrm{BM}(\mu  \ge 0)
      $
      &
      $\displaystyle
      \mathrm{BM}(\mu)
      $
      \\
      
      \noalign{\vskip 8pt} 
      $\tanh(\alpha, \beta)$ 
      &
      $\displaystyle
      \mathrm{BM}(\mu <0)
      $
      &
      $\displaystyle
       \alpha + \frac{-(\mu +\alpha ) e^{2 a \mu -\mu  x-\alpha  x +\frac{1}{2} \left(\alpha ^2-\mu ^2\right)t}+2 \alpha \frac{\left(1-e^{2 a \mu }\right)}{\left(1-e^{2 a \alpha }\right)} e^{2 \alpha (a-x)}}{e^{2 a \mu -\mu  x-\alpha  x +\frac{1}{2} \left(\alpha ^2-\mu ^2\right)t} +\frac{\left(1-e^{2 a \mu }\right)}{\left(1-e^{2 a \alpha }\right)} \left(1-e^{2 \alpha (a-x)}\right) }
      $
      \\
      
      \noalign{\vskip 8pt} 
      $\tanh(\alpha, \beta)$ 
      &
      $\tanh(\gamma, \delta)$
      &
      $\displaystyle
      {\frac{\gamma  \left(e^{2 (a \gamma +\delta )}+1\right) e^{\frac{1}{2} (\alpha^2 -\gamma^2 ) t+\gamma  x}-\frac{\alpha  \left(e^{2 a \gamma }-1\right) e^{\alpha  x} \left(e^{2 \alpha  (a-x)}+1\right)}{e^{2 a \alpha }-1}}{\left(e^{2 (a \gamma +\delta )}+1\right) e^{\frac{1}{2} (\alpha^2 -\gamma^2 ) t+\gamma  x}+\frac{\left(e^{2 a \gamma }-1\right) e^{\alpha  x} \left(e^{2 \alpha  (a-x)}-1\right)}{e^{2 a \alpha }-1}}  }
      $
      \\
      
      \noalign{\vskip 8pt} 
      $\tanh(\alpha, \beta)$ 
      &
      $\tanh(\alpha, \delta) $
      &
      $\displaystyle
      \tanh(\alpha, \delta)
      $
      \\
            
      \noalign{\vskip 8pt} 
      $\displaystyle
      \mathrm{BM}(\mu \ge 0) $
      &
      $\tanh(\alpha, \beta) $
      &
      $\displaystyle
      \frac{
        a\alpha\,e^{\alpha x+\beta}\cosh(a\alpha+\beta)\;-\;\sinh(a\alpha)\,e^{\frac{\alpha^{2}t}{2}}
        }{
        a\,e^{\alpha x+\beta}\cosh(a\alpha+\beta)\;+\;\sinh(a\alpha)\,e^{\frac{\alpha^{2}t}{2}}(a-x)
        }
      $
      \\
            
      \noalign{\vskip 8pt} 
      $\displaystyle
      \mathrm{BM}(\mu \ge 0) $
      &
      $\mathrm{taboo}(b)$
      &
      $\displaystyle
      \mathrm{taboo}(b)
      $
      \\
                  
      \noalign{\vskip 8pt} 
      $\displaystyle
      \mathrm{BM}(\mu < 0) $
      &
      $\tanh(\alpha, \beta) $
      &
      $\displaystyle
      \frac{\frac{2a\,e^{a\alpha+\beta}\cosh(a\alpha+\beta)\,
e^{-\tfrac{1}{2}(\alpha^2-\mu^2)t}\,e^{\alpha x}}{a-x}
\left(\alpha+\frac{1}{a-x}\right)
\;-\;
2\mu\,e^{a\alpha}\,\frac{\sinh(a\alpha)}{\sinh(a\mu)}\,
\cosh\!\bigl(\mu(a-x)\bigr)}
{\frac{2a\,e^{a\alpha+\beta}\cosh(a\alpha+\beta)\,
e^{-\tfrac{1}{2}(\alpha^2-\mu^2)t}\,e^{\alpha x}}{a-x}
\;+\;
2e^{a\alpha}\,\frac{\sinh(a\alpha)}{\sinh(a\mu)}\,
\sinh\!\bigl(\mu(a-x)\bigr)}
      $
      \\
      
      \noalign{\vskip 8pt} 
      $\displaystyle
      \mathrm{BM}(\mu < 0) $
      &
      $\mathrm{taboo}(b)$
      &
      $\displaystyle
       \mu + \frac{-\mu  (b-a) e^{\frac{\mu ^2 t}{2}-\mu  x}+\frac{2 a \mu  e^{2 \mu  (a-x)}}{\left(1-e^{2 a \mu }\right)}}{{(b-a) e^{\frac{\mu ^2 t}{2}-\mu  x}}+\frac{a \left(1-e^{2 \mu  (a-x)}\right)}{ \left(1-e^{2 a \mu }\right)}} 
      $
      \\
                  
      \noalign{\vskip 8pt} 
      $\displaystyle
      \mathrm{taboo}(b)
      $ 
      &
      $\displaystyle
      \mathrm{BM}(\mu  \ge 0)
      $
      &
      $\displaystyle
      \mathrm{BM}(\mu)
      $
      \\
                  
      \noalign{\vskip 8pt} 
      $\displaystyle
      \mathrm{taboo}(b)
      $ 
      &
      $\displaystyle
      \mathrm{BM}(\mu  < 0)
      $
      &
      $\displaystyle
      \frac{2\sinh(a\mu)\,e^{\frac{\mu^2 t}{2}+\mu(x-a)} - a\mu}{a - 2(a-x)\sinh(a\mu)\,e^{\frac{\mu^2 t}{2}+\mu(x-a)}}
      $
      \\
    \end{tabular}
  \end{ruledtabular}
\end{table}

\newpage
\section{Characterization of returnability under first-passage-time conditioning}
\label{appendix_inverse}
In his seminal 1931 paper \emph{``On the Reversal of the Laws of Nature''}, Schr\"odinger showed that imposing endpoint constraints on a diffusion leads to a time-reversed dynamics through a suitably tilted probability measure~\cite{Schrodinger,CommentSchrodinger}. 
In the present framework, we address a closely related question: namely, whether distinct diffusion processes can be obtained from one another by conditioning the first on the first-passage-time distribution of the second, and conversely?

\noindent Structural symmetries---including the emergence of reversible cycles under successive first-passage-time conditionings---can be observed in Tab.~\ref{table_densities}. 
Table~\ref{table_densities} indeed shows that such reversibility arises in several families of drifts. 
For instance, conditioning a $\tanh(\alpha,\beta)$ process on the FPT density of a $\tanh(\alpha,\delta)$ process yields a $\tanh(\alpha,\delta)$ drift; conditioning this latter process on the $\tanh(\alpha,\beta)$ FPT density brings one back to the original dynamics. 
A similar back-and-forth behavior occurs for Brownian motion with a positive constant drift~\cite{ref_Monthus_Mazzolo}, whereas no such reversibility is observed for negative constant drifts. 
More importantly, such reciprocal behavior is not restricted to diffusions of the same type: Brownian motion with a constant positive drift and the taboo process form a reciprocal pair, since each can be obtained by conditioning the other on its first-passage-time distribution.

Let $X(t)$ be a diffusion process with drift $\mu(x,t)$, and assume that $X^{*}(t)$ is obtained from $X(t)$ by conditioning on a first-passage-time density, so that the drift of the conditioned process reads
\begin{equation}
\mu^{*}(x,t) = \mu(x,t) + \partial_x \log Q^{[\mu\to\mu^{*}]}(x,t) \, ,
\end{equation}
where the quantity $Q^{[\mu\to\mu^{*}]}(x,t)$, which depends on both the imposed FPT density and the associated survival condition, has been given in the main text. For such a conditioned process, the corresponding probability density reads
\begin{equation}
\rho^*(x,t) = Q^{[\mu\to\mu^{*}]}(x,t)\,\rho(x,t) \, .
\end{equation}
A second conditioning of $X^{*}(t)$, based on a (possibly different) FPT law, produces another diffusion whose drift depends on the corresponding function $Q^{[\mu^{*}\to\mu]}(x,t)$.

Observe that $Q^{[\mu\to\mu^{*}]}(x,t) >0$. A \emph{necessary and sufficient condition} for this second conditioning to recover the original dynamics—namely, to produce a process with drift $\mu(x,t)$ and hence probability density $\rho(x,t)$—is that the two functions $Q^{[\mu\to\mu^{*}]}(x,t)$ and $Q^{[\mu^{*}\to\mu]}(x,t)$ are mutual inverses:
\begin{equation}
Q^{[\mu^{*}\to\mu]}(x,t)
=
\frac{1}{Q^{[\mu\to\mu^{*}]}(x,t)} .
\label{Q_inverse}
\end{equation}
The proof is straightforward. Indeed, the density of the twice-conditioned process is
\begin{equation}
\rho(x,t)
=
Q^{[\mu^{*}\to\mu]}(x,t)\,\rho^{*}(x,t)
=
Q^{[\mu^{*}\to\mu]}(x,t)\,Q^{[\mu\to\mu^{*}]}(x,t)\,\rho(x,t) \, ,
\end{equation}
which directly implies Eq.~\eqref{Q_inverse}.

\smallskip
This result can be explicitly checked on the example of a Brownian motion with positive constant drift and the taboo process, which constitute a reciprocal pair under first-passage-time conditioning. Indeed, one can observe that the function $Q^{[\mu \to \mathrm{taboo}]}(x,t)$ given by Eq.~\eqref{Qmupostotaboo} and the function $Q^{[\mathrm{taboo} \to \mu]}(x,t)$ given by Eq.~\eqref{QtimeinfinitytabootoBM} are exact inverses of one another.

\smallskip
The previous result is useful in that it provides a criterion for determining when a return is possible. However, it remains incomplete, as it does not specify which classes of processes or which forms of conditioning actually satisfy this criterion.
While the result allows one to assess the possibility of a return once the functions $Q(x,t)$ are known, a more detailed analysis is required to fully identify and characterize the processes and conditionings for which such reversibility can occur. This broader classification lies beyond the scope of the present work and is left for future study.

\section*{AUTHOR DECLARATIONS}
\subsection*{Conflict of Interest}
The authors have no conflicts to disclose.

\subsection*{Author Contributions}
{\bf{Kacim Fran\c{c}ois-\'{E}lie}}: Formal analysis (equal); Investigation (equal); Writing – original draft (equal).
{\bf{Alain Mazzolo}}: Formal analysis (equal); Investigation (equal); Writing – original draft (equal).
 
\section*{DATA AVAILABILITY}
Data sharing is not applicable to this article as no new data were created or analyzed in this study.


\end{document}